%% file: 2024-PAPER-KIS_selfBalacing.tex
\documentclass[reprint,linenumber, amsmath amssymb, titlepage, aps, pra, superscriptaddress, raggedbottom, nofootinbib, floatfix, citeautoscript]{revtex4-2}
\usepackage{silence}
\usepackage[utf8]{inputenc}
\usepackage[english]{babel}
\usepackage{hyperref}
\usepackage{bm}
\usepackage{hyperref}
\usepackage{float}
\usepackage[newcommands]{ragged2e}
\usepackage{physics}
\usepackage{tabularx}
\usepackage{mathtools}
\usepackage[raggedright]{titlesec}
\usepackage[toc,page,header]{appendix}
\usepackage{minitoc}
\usepackage{nicefrac}
\usepackage{sidecap}
\usepackage{multibib}
\usepackage[dvipsnames]{xcolor}
\usepackage{booktabs}
\usepackage{float}
\usepackage{graphicx}
\usepackage[capitalise]{cleveref}
\usepackage[version=4]{mhchem}
\usepackage{tabularx}
\usepackage{ulem}
\usepackage{enumitem}

\sidecaptionvpos{figure}{c}
\usepackage[per-mode=power, 
            exponent-product = \times, 
            inter-unit-product = \ensuremath{{\cdot}}, 
            tight-spacing = true,
            number-unit-product=~,
            parse-numbers=false,]{siunitx}
\AtBeginDocument{\RenewCommandCopy\qty\SI}
\ExplSyntaxOn
\msg_redirect_name:nnn { siunitx } { physics-pkg } { none }
\ExplSyntaxOff
\hypersetup{colorlinks=true,  linkcolor=blue, 
            citecolor=blue, filecolor=blue,  
            urlcolor=blue}
\usepackage{caption}
\titleformat*{\section}{\bfseries}
\titleformat*{\subsection}{\bfseries}
\titlespacing{\section}{0em}{1em}{0.25em}
\renewcommand{\textcite}[1]{\citenum{#1}}
\newcolumntype{Y}{>{\centering\arraybackslash}X}
\Crefformat{appendix}{Supplementary Information #2#1#3}
\let\origaddcontentsline\addcontentsline

\newcommand{\enabletocentries}{%
  \let\addcontentsline\origaddcontentsline
}
\DeclareSIUnit[qualifier-mode = combine]{\dBm}{\deci\bel\of{m}}

\begin{document}
\newcommand{\mytitle}{Dissipative Kerr Soliton Self-Balancing from Kerr-Induced Synchronization}
\title{\mytitle}
\author{Pradyoth Shandilya}
\affiliation{University of Maryland Baltimore County, Baltimore, MD, USA}
\author{Kartik Srinivasan}%
\affiliation{Joint Quantum Institute, NIST/University of Maryland, College Park, USA}
\affiliation{Microsystems and Nanotechnology Division, National Institute of Standards and Technology, Gaithersburg, USA}
\author{Curtis Menyuk}%
\affiliation{University of Maryland Baltimore County, Baltimore, MD, USA}
\author{Gr\'egory Moille}%
\email{gmoille@umd.edu}
\affiliation{Joint Quantum Institute, NIST/University of Maryland, College Park, USA}
\affiliation{Microsystems and Nanotechnology Division, National Institute of Standards and Technology, Gaithersburg, USA}
\date{\today}

\begin{abstract}    
    Integrated frequency comb sources are a key enabling technology for frequency metrology applications. Their on-chip integration promises to bring metrology capacity outside of the lab, particularly since they can operate at low continuous-wave pump laser power in the dissipative Kerr soliton (DKS) regime. Yet, such small foot-print and low power comes at a cost: higher noise and overall lower comb power. In particular, this translates to highly challenging detection and locking of the carrier-envelope offset, necessary for complete stabilization of the comb. %, remains elusive because of the overall low power of each comb teeth. %
    Recently, Kerr-induced synchronization (KIS) of a DKS to a reference laser has been demonstrated as a tool for passive all-optical stabilization of DKS microcombs, with fundamental modification to the DKS and microcomb properties. Here, we demonstrate that the combination of additional power from the reference laser (now part of the DKS) and the KIS phase locking that pins the repetition rate together fundamentally alter the DKS, forcing an energy redistribution to maintain its center of mass. %
        We demonstrate this self-balancing effect theoretically, which in a pure quadratic dispersion resonator leads to reference-dependent recoil. %
        With higher-order dispersion through which the DKS yields phase-matched dispersive waves (DWs), we demonstrate that self-balancing increases the DW radiation, experimentally showing a \qty{22}{\dB} increase of comb teeth at \qty{780}{\nm} in an octave-spanning microcomb for efficient deployable carrier-envelope offset detection. 
\end{abstract}

\maketitle
\section{Introduction}

\begin{figure*}[t]
    \centering
    \includegraphics{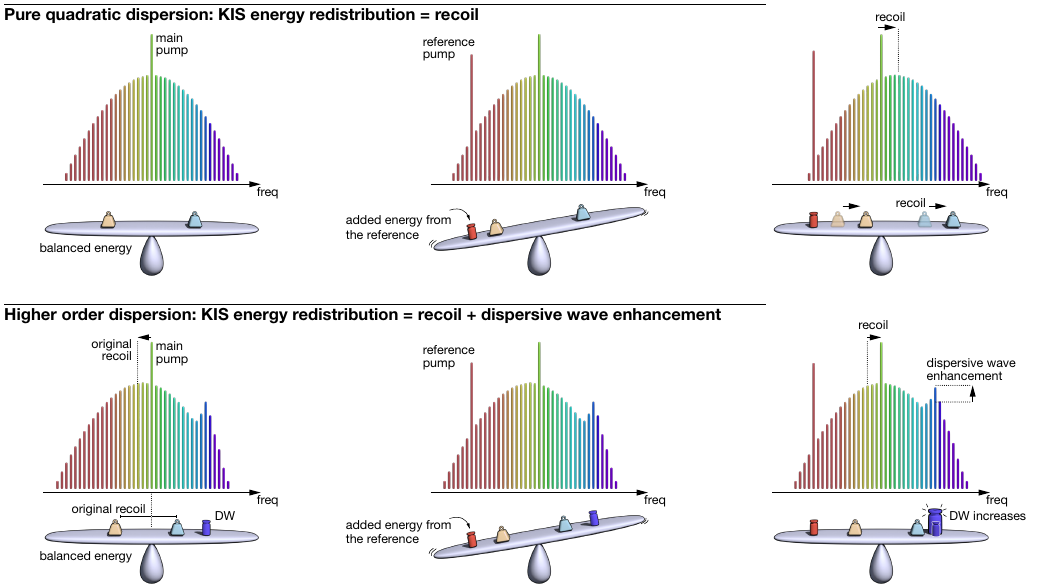}
    \caption{\label{fig:1}\textbf{Soliton self-balancing mechanism through spectral energy redistribution under Kerr-induced synchronization---}
    \textbf{(a--c) Pure quadratic dispersion case.} %
    \textbf{a} The DKS frequency comb centers on the main pump. %
    The equivalent lever diagram represents conservation of the center of mass, which in the single-pump case corresponds to equal weight at equal distance from the center (\textit{i.e.}, the main pump). %
    \textbf{b} In the KIS regime, however, the reference pump becomes part of the DKS and adds asymmetric energy, creating a nonphysical imbalance. %
    \textbf{c} Because the repetition rate is pinned by the main and reference pumps, the DKS cannot compensate for the energy imbalance through temporal drift. %
    The sole degree of freedom available to the DKS is to shift its center of mass, resulting in recoil. %
    \textbf{(d--f) Higher-order dispersion case.} %
    \textbf{d} If the resonator exhibits at least a tertiary dispersion term, the DKS becomes phase-matched to at least one azimuthal mode of the cavity, leading to dispersive wave emission and creating a natural imbalance, hence an intrinsic recoil for center-of-mass conservation. %
    \textbf{e} Similarly, in the KIS regime the reference pump adds an energy imbalance. %
    \textbf{f} The DKS now possesses another degree of freedom to compensate for the reference energy. %
    While it still compensates via recoil, it also increases its dispersive wave emission, enhancing the dispersive wave mode away from the reference mode. %
    DW: dispersive wave. 
        }  
\end{figure*}

\begin{figure}[t]
    \centering
    \includegraphics{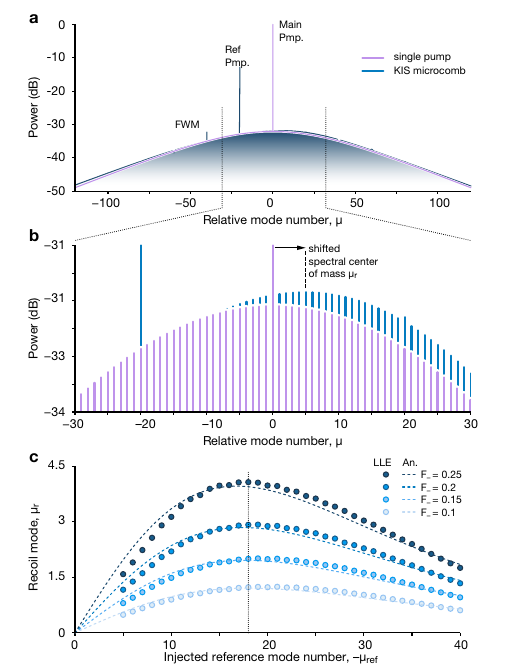}
    \caption{\label{fig:2}\textbf{Self-balancing in a pure quadratic integrated dispersion microresonator. }
    \textbf{(a)} Simulated spectra with (blue) and without (purple) KIS. FWM: four-wave mixing. 
    \textbf{(b)} Spectral recoil that occurs due to KIS. When the reference pump is used to synchronize the DKS on one side of the spectrum, a spectral recoil occurs and the spectral center of mass is shifted in the opposite direction. 
    \textbf{(c)} Comparison of simulated and analytically calculated spectral recoil at varying reference pump mode numbers and amplitudes. The simulated spectral recoil (circular markers) follow the analytically calculated $\sech(\cdot)\tanh(\cdot)$ trend (dotted lines). The maximal spectral recoil is obtained at the same $\mu_\mathrm{ref}$ for all reference pump amplitudes.}  
\end{figure}

Optical frequency combs based on dissipative Kerr solitons (DKSs) in microresonators have emerged as low power~\cite{SternNature2018}, mass-scale fabrication compatible~\cite{LiuNatCommun2021,OuOpt.Lett.OL2025}, and versatile coherent light sources for metrology~\cite{SpencerNature2018,NewmanOptica2019}, communications~\cite{CorcoranNat.Photon.2025}, spectroscopy~\cite{DuttSci.Adv.2018}, distance ranging~\cite{RiemensbergerNature2020}, and other applications. % 
However, like table-top counterparts, these microcombs require complete stabilization for metrological applications, typically involving detection and locking of their carrier-envelope offset (CEO) frequency. %
Yet, their low pump power and small footprint directly translate into low power per comb tooth and larger thermo-optic noise~\cite{HuangPhys.Rev.A2019}, leading to poor signal-to-noise ratio~\cite{DrakeNat.Photonics2020} in nonlinear $f-2f$ interferometry for CEO detection~\cite{TelleApplPhysB1999a}. %
Overcoming this limitation is essential for advancing field-deployable microcomb systems that combine broad spectral coverage with high coherence. %
A promising pathway towards such control is Kerr-induced synchronization (KIS)~\cite{MoilleNature2023}, where a second laser injected into the same microresonator housing the DKS phase synchronizes the soliton~\cite{MoillePhys.Rev.Lett.2025b}, enabling passive stabilization and capture of a comb tooth. %
Thus, KIS provides a mechanism for passive two-point external control of the DKS microcomb, to achieve ultra-low noise operation~\cite{SunNat.Photon.2025}. %
This ultra-low noise performance results from a fundamental modification of the DKS dynamics under KIS, breaking its symmetry to bypass intra-cavity noise and alter tooth-to-tooth noise propagation~\cite{MoilleOptica2025}. %
\\
\indent In this work, we demonstrate that, by constraining the soliton's group velocity through the fixed frequency difference between the two optical pumps, KIS introduces a new degree of freedom in controlling the intracavity energy flow and spectral symmetry of the soliton state. %
We reveal that under KIS, the DKS undergoes a self-balancing process that conserves its spectral center of mass despite the energy imbalance introduced by the reference pump. %
Through a perturbative analysis of the multi-pump Lugiato-Lefever equation (MLLE)~\cite{TaheriEur.Phys.J.D2017}, we show that the soliton experiences a compensating spectral recoil directed opposite to the reference frequency (\cref{fig:1}a-c), effectively redistributing energy within the comb to maintain its equilibrium. %
Importantly, this mechanism, which we term soliton self-balancing, has no analogue in conventional single-pumped systems, where recoil is generally associated with changes in the repetition rate rather than spectral re-equilibration. %
Moving beyond the case of pure quadratic dispersion, for systems with higher-order dispersion this self-balancing behavior can transfer significant energy into a dispersive wave (DW) opposite to the reference pump in the optical spectrum, resulting in a power enhancement of high-frequency DWs (\cref{fig:1}d-f). %
Experimentally, we confirm this prediction by observing a \qty{22}{\dB} increase in the high-frequency DW of an octave-spanning microcomb when operated in the KIS regime. %
This enhancement arises not from conventional four-wave mixing between the main pump and reference laser, but from the soliton's intrinsic drive to preserve its spectral center of mass under dual optical pinning. %
The result effectively ``amplifies'' the weaker DW without modifying the repetition rate, thereby facilitating CEO detection in otherwise low-power microcombs. The concept of soliton self-balancing extends the understanding of how nonlinear optical cavities redistribute energy under multiple driving fields. By linking spectral recoil, DW dynamics, and KIS locking, our findings reveal a new self-organizing behavior of cavity solitons. Beyond providing a route to more efficient CEO detection, self-balancing opens opportunities for engineered spectral control and broadband enhancement in chip-scale frequency combs.

\section{Self-balancing in purely quadratic dispersion resonators}

In the simplest system where the resonator exhibits only quadratic dispersion, a DKS follows a perfect $\sech$-squared envelope without recoil. %
Consequently, the output state produces a frequency comb with repetition rate $\nu_\mathrm{rep}^0$ that is exactly symmetric about the primary pump. %
Equivalently, its spectral center of mass coincides with the pumped mode (\cref{fig:1}a). %
KIS of the DKS from a second injected laser on either side of the main pump causes the in-phase component of the intracavity reference field to merge with the DKS. %
Hence, it leads to a displacement of the spectral center of mass (SCM) away from the reference laser azimuthal mode (\cref{fig:1}b). %
However, due to the optical trapping of the DKS inside the potential created by the modulation between the main and reference pumps, its group and phase velocities are now entirely pinned by these two laser frequencies such that $\nu_\mathrm{rep}^\textrm{KIS} = (\nu_\mathrm{pmp}-\nu_\mathrm{ref})/M$, where $\nu_\mathrm{rep}^\textrm{KIS}$ is the disciplined soliton repetition rate, $\nu_\mathrm{pmp}$ and $\nu_\mathrm{ref}$ are the main and reference pump frequencies, and $M$ is the number of modes separating the two pumps. 
% However, when the laser is placed at or sufficiently near an existing comb line, the soliton can enter the KIS regime where the repetition rate is dictated solely by the frequencies of the two pumps and the number of modes separating them, given by $\nu_\mathrm{rep} = (\nu_\mathrm{pmp}-\nu_\mathrm{ref})/M$, where $\nu_\mathrm{rep}$ is the soliton repetition rate, $\nu_\mathrm{pmp}$ and $\nu_\mathrm{ref}$ are the main pump and reference (injected) pump frequencies, and $M$ is the number of modes separating the two pumps. %
When at the center of the KIS window, namely $\nu_\mathrm{ref} =\nu_\mathrm{pmp} + M\nu_\mathrm{rep}^0$, where the reference pump frequency exactly matches the frequency of one of the comb lines before KIS, there is no change in the repetition rate before and after KIS ($\nu_\mathrm{rep}^0 = \nu_\mathrm{rep}^\textrm{KIS}$), even though more power has been injected into the microresonator asymmetrically with respect to the main pump frequency. %
Consequently, to preserve its repetition rate under KIS, the soliton must re-establish the original SCM . %
With only quadratic dispersion available, the sole route to do so is via a compensating spectral recoil away from the reference pump frequency (\cref{fig:1}c). %
Moving next to the case when higher-order dispersion is present, the possibility of phase matching between the cavity resonances and the soliton exists, leading to the creation of DW(s), which will cause a recoil in the single-pump case (\cref{fig:1}d). %
Therefore, addition of the reference pump leads to the same energy imbalance as in the pure quadratic dispersion case (\cref{fig:1}e), introducing yet another degree of freedom to enable self-balancing. %
The dispersive radiation from the soliton can increase, leading to an enhancement of the DW opposite to the reference in the optical spectrum (\cref{fig:1}f). %
We study the latter case in the next section; in this section, we begin our study of soliton self-balancing in the simplest possible system to gain physical insight: a microresonator exhibiting purely quadratic dispersion.
\\
\indent The impact of the reference pump on the DKS spectral energy distribution can be analyzed using standard soliton perturbation theory~\cite{KaupPhysRevA1990,Haus1990,GeorgesOptFibTech1995}, which has been successfully used to study soliton dynamics in microresonators~\cite{Mizrahi2024,Leshem2025}. Soliton dynamics in microresonators pumped by two lasers is described using the loss-normalized multi-pump Lugiato-Lefever equation (MLLE)~\cite{TaheriEur.Phys.J.D2017,MoilleOptica2025}
\begin{equation}\label{eq:mlle}
    \begin{split}
        \frac{\partial a}{\partial t} = &-\left( 1 + i\alpha_{\rm pmp} \right)a + i\sum_\mu D_{\rm int}(\mu)A(\mu) e^{i\mu\theta} + i|a|^2a \\
                                        &+ F_{\rm pmp} + F_{\rm ref}e^{i(\alpha_{\rm ref} - \alpha_{\rm pmp} + D_{\rm int}(\mu_{\rm ref}))t + i\mu_{\rm ref}\theta},
    \end{split}
\end{equation}
where $a$ is the intracavity field amplitude, $A$ is its Fourier transform, $t$ is time, $\theta$ is the azimuthal coordinate, $\alpha_{\rm pmp}$ ($\alpha_{\rm ref}$) is the detuning of the primary (reference) pump with corresponding amplitudes $F_{\rm pmp}$ ($F_{\rm ref}$), $D_{\rm int}$ is the integrated dispersion as a function of the relative mode number $\mu$, and $\mu_s$ is the relative mode number of the reference pump. Considering only up to second-order dispersion, \cref{eq:mlle} reduces to 
\begin{equation}\label{eq:mlle_D2}
    \begin{split}
                \frac{\partial a}{\partial t} = &-\left( 1 + i\alpha_{\rm pmp} \right)a + iD_2 \frac{\partial^2 a}{\partial \theta^2} + i|a|^2a \\
                                        &+ F_{\rm pmp} + F_{\rm ref}e^{i(\alpha_{\rm ref} - \alpha_{\rm pmp} + D_{\rm int}(\mu_{\rm ref}))t + i\mu_{\rm ref}\theta},
    \end{split}
\end{equation}
where $D_2$ is the second-order dispersion coefficient. Normalizing \cref{eq:mlle_D2} with respect to the detuning (see Supplementary Material), we obtain
\begin{equation}\label{eq:mlle_det_normalized}
    \begin{split}
        \frac{\partial \psi}{\partial t} = &-\left(\delta + \frac{i}{2} \right)\psi + \frac{i}{2}\frac{\partial^2 \psi}{\partial x^2} + i|\psi|^2\psi \\
        &- \frac{i}{2}h_{\rm pmp} - \frac{i}{2}h_{\rm ref}e^{i\mu_s x} - \sigma \frac{\partial \psi}{\partial x},
    \end{split}
\end{equation}
where $\psi(x,t)$ is the normalized intracavity field amplitude defined in the normalized azimuthal coordinate $x$, $\delta$ is the normalized loss, $h_{\rm pmp}$ ($h_{\rm ref}$) is the normalized main (reference) pump amplitude, and $\sigma$ is the normalized detuning of the reference pump from its closest comb line frequency. We can treat the loss and gain terms perturbatively and rewrite \cref{eq:mlle_det_normalized} as
\begin{equation}\label{eq:mlle_perturbed}
    \begin{split}
        \frac{\partial \psi}{\partial t} = -\frac{i}{2}\psi + \frac{i}{2}\frac{\partial^2 \psi}{\partial x^2} + i|\psi|^2\psi + \epsilon\mathcal{P}(x,t),
    \end{split}
\end{equation}
where $\epsilon\mathcal{P}(\psi,x,t) = -\delta\psi + \sigma{\partial \psi}/{\partial x} - (i/2)[h_{\rm pmp} +h_{\rm ref}e^{i\mu_s x}]$ are the terms we will treat perturbatively, and the $\epsilon$ coefficient is used to indicate that the terms are small. The MLLE as written in \cref{eq:mlle_perturbed} is convenient for theoretical analysis because it is of the form of a perturbed nonlinear Schr\"odinger equation (NLSE), which enables the use of standard soliton perturbation theory to study the impact of the reference pump on the soliton. We therefore assume that the solution is of the form
\begin{equation}\label{eq:ansatz}
    \psi_s(x,t) = A\sech[A(x-x_0)]e^{i\mu_r(x-x_0) - i\phi},
\end{equation}
where the parameters $p_j = {A,x_0,\mu_r, \phi}$ are the soliton amplitude, central position, central frequency (spectral recoil), and central phase, respectively. We assume that only these four parameters are subject to variation due to the presence of a perturbation. With this assumption, we employ soliton perturbation theory to calculate how $x_0$ changes with time in the presence of the loss and gain perturbative terms (see Supplementary Material), and we obtain
\begin{equation}\label{eq:perturbation_drift}
    \frac{d x_0}{dt} = h_\mathrm{DKS}^2\mu_r - \sigma + \frac{h_{\rm ref} \pi^2}{4A^2}\sech\left( \frac{\mu_s \pi}{2A} \right)\tanh\left( \frac{\mu_s \pi}{2A} \right),
\end{equation}
where $h_\mathrm{DKS} =\sqrt{({\pi}/{A})\int_{-\infty}^{\infty}|\Psi_s(\mu)|^2 d\mu}$ is the normalized soliton energy, ${x \leftrightarrow \mu}$ are Fourier transform pairs, and ${\Psi_s(\mu) = \int_{-\infty}^{\infty}\psi_s(x)e^{-i\mu x}dx}$ is the Fourier transform of $\psi_s(x)$. When the soliton is in the KIS regime, \cref{eq:mlle_perturbed} has stationary solutions. Therefore, in KIS, $dx_0/dt = 0$, and \cref{eq:perturbation_drift} becomes
\begin{equation}\label{eq:self_balancing_D2}
\begin{split}
    h_\mathrm{DKS}^2\mu_r &=  -\frac{h_{\rm ref} \pi^2}{4A^2}\sech\left( \frac{\mu_s \pi}{2A} \right)\tanh\left( \frac{\mu_s \pi}{2A} \right) + \sigma \\
    h_\mathrm{DKS}^2\mu_r &=  -h_\mathrm{KIS}^2\tanh\left( \frac{\mu_s \pi}{2A} \right) + \sigma.
\end{split}
\end{equation}
We note $h_\mathrm{KIS}^2 = ({\pi^2}/{4A^2})h_{\rm ref}  \sech(\mu_s \pi/2A)$, and $h_\mathrm{KIS}$ has been previously defined as the KIS energy~\cite{MoilleNature2023,WildiAPLPhotonics2023} equal to the normalized geometric mean of the reference pump energy $h_\mathrm{ref}$ and the power of the comb tooth where the synchronization occurs $h(\mu_s)= \sech(\mu_s \pi/2A)$.
We first consider the case where the reference pump is injected at a frequency exactly equal to the frequency of one of the comb lines, i.e., when $\sigma=0$. In this case, \cref{eq:self_balancing_D2} shows that there is indeed a spectral recoil ($\mu_r\neq0$), as shown in \cref{fig:2}(b), even though there is no change in the DKS repetition rate. This results in a behvaior similar to the all-optical phase trapping through cavity modulation of a table-top fiber comb, which could be used to counteract the spectral recoil arising from the Raman effect~\cite{Englebert2024}. Our results present the main advantage of KIS-induced self-balancing to be an all-optical phenomeonon and therefore compatible with any cavity FSR, and hence with an integrated DKS platform. This result is significant by itself because it shows that KIS exhibits a spectral recoil without a change in its repetition rate, unlike systems with Raman-induced spectral recoil \cite{MilianPhysRevA2015,YiOptLett2016} or high-order dispersion \cite{CherenkovPhysRevA2017,akhmediev_cherenkov_1995}, where a spectral recoil is necessarily associated with a change in the repetition rate. Furthermore, \cref{eq:self_balancing_D2} has the form of a conservation of torque equation because it can be written in the form ${m_1 l_1 = m_2l_2}$ where $m_1 = h_\mathrm{DKS}^2$ and $m_2 = h_\mathrm{KIS}^2$ represent mass-like quantities and ${l_1=\mu_r}$ and ${l_2 = -\tanh(\mu_s\pi/2A)}$ represent length-like quantities. Therefore, \cref{eq:self_balancing_D2} provides the magnitude of the spectral recoil needed for the DKS to maintain its repetition rate. The right-hand side of \cref{eq:self_balancing_D2} has a distinctive $\sech(\cdot)\tanh(\cdot)$ form, implying that for a given reference pump amplitude, there exists an optimal mode number $\mu_s$ at which the reference pump can be injected to maximize the recoil, and that this optimal mode number is independent of the reference pump amplitude. To verify this result, we simulate \cref{eq:mlle_D2} fixing all parameters except $\mu_{\rm ref}$ and varying $\mu_{\rm ref}$ for different values of $F_{\rm ref}$. The simulated spectral recoil [\cref{fig:2}(c)] follows the $\sech(\cdot)\tanh(\cdot)$ trend predicted in \cref{eq:self_balancing_D2}, and the maximum spectral recoil occurs at the same reference pump mode number for all reference pump amplitudes. The numerical simulations are consistent with the results of perturbation theory that demonstrate the presence of self-balancing.
\section{Self-balancing effect in presence of a dispersive wave}

\begin{figure*}[t]
    \centering
    \includegraphics{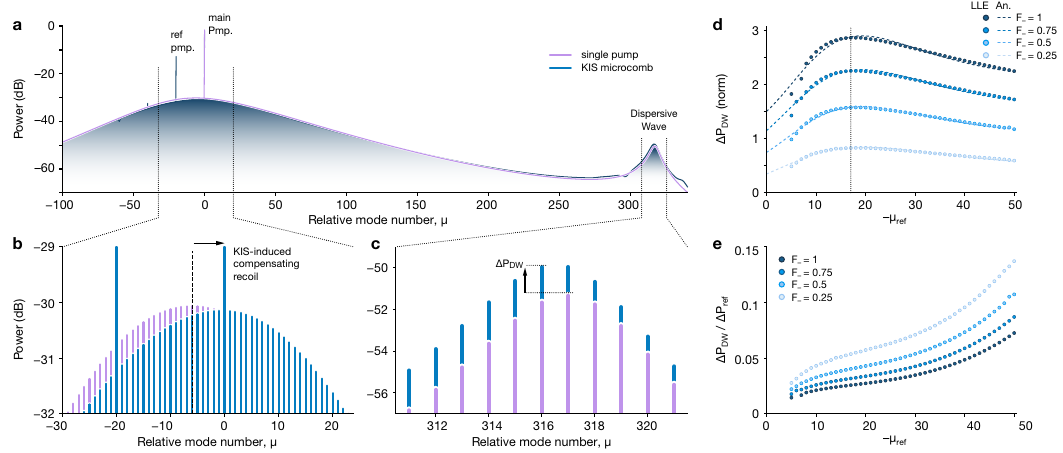}
    \caption{\label{fig:3} \textbf{Demonstration of self-balancing effect in the presence of a single DW. }
    \textbf{(a)} Comparison of the frequency comb spectra with (blue) and without (purple) KIS. In the absence of KIS, the peak of the spectrum is offset from $\mu=0$ due to a spectral recoil resulting from the presence of a DW. When a reference pump is injected at the frequency of the comb line at $\mu_{\rm ref}=-20$, the soliton enters the KIS regime and experiences a spectral recoil in a manner identical to what was shown in the pure-quadratic case. \textbf{(b)} The KIS-induced spectral recoil opposes the recoil due to the DW and cancels each other out, leaving the resulting spectrum nearly symmetric around $\mu=0$. \textbf{(c)} Apart from the spectral recoil, the self-balancing effect leads to a significant increase in the power of the comb lines around the DW mode. 
    \textbf{(d)} Change in the DW power as a function of the mode at which the reference pump is injected, for different reference pump powers. The circular points show results obtained from the MLLE and dotted lines are fits to the analytical expression obtained in \cref{eq:delta_dw_power}. A clear peak exists at the same $\mu_{\rm ref}$ regardless of the power of the reference pump and the DW power decreases monotonically away from this peak, clearly demonstrating that the DW enhancement in the presence of the reference pump is not due to simple four-wave mixing between the main and reference pumps. 
    \textbf{(e)} The ratio of the DW power and the reference pump power as a function of the reference pump mode number, indicating the efficiency of DW enhancement due to the self-balancing effect. The ratio increases monotonically with $\mu_{\rm ref}$ in agreement with \cref{eq:self_balancing_D3}.} 
\end{figure*}

\begin{figure*}[t]
    \centering
    \includegraphics{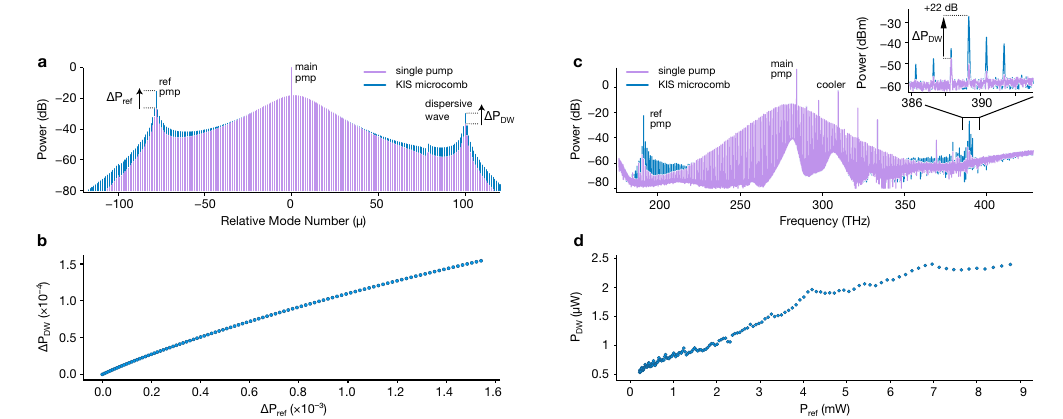}
    \caption{\label{fig:4} \textbf{Numerical and experimental demonstration of DW power enhancement due to self-balancing. }
    \textbf{(a)} Numerically calculated spectra with (blue) and without (purple) KIS, demonstrating that the injection of the reference pump at the low-frequency DW leads to a proportional increase in the high-frequency DW power $\Delta P_\mathrm{DW}$ on the other side of the spectrum. 
    \textbf{(b)} The change in the high-frequency DW power is linearly proportional to the change in the comb tooth power where the reference pump is injected, as predicted analytically. 
    \textbf{(c)} Experimental spectra with (blue) and without (purple) KIS showing that the injection of the reference pump at the low-frequency DW leads to a \qty{\approx22}{\dB} enhancement of the high-frequency DW. 
    \textbf{(d)} Experimental measurement of the high frequency DW power against reference power. The uncertainty are within the marker size. At low reference pump power, the DW power increases linearly with the reference pump power, as predicted analytically and verified numerically in (c).} 
\end{figure*}

In the previous section, we observed analytically and numerically through MLLE simulations that the injection of a reference pump causes the soliton to undergo a spectral recoil to conserve its spectral center of mass without changing the repetition rate, lending credence to the notion of self-balancing. Ultimately, however, we are interested in showing that the self-balancing effect can lead to an increase in DW power as a way to conserve the spectral center of mass. Additionally, we want to show that an increase in DW power is indeed due to self-balancing and not through simple four-wave mixing between the main and reference pumps. Since DWs were not present in the simple system with a pure quadratic integrated dispersion that we considered in the previous section, we modify the integrated dispersion to include a small but non-negligible third-order dispersion term, which leads to the formation of a DW at a position relatively far away from the center of the soliton spectrum. Including this term modifies \cref{eq:mlle_D2}, which now becomes
\begin{equation}\label{eq:mlle_D3}
    \begin{split}
        \frac{\partial a}{\partial t} = &-\left(1 + i\alpha_{\rm pmp} \right)a + iD_2 \frac{\partial^2 a}{\partial \theta^2} + D_3 \frac{\partial^3 a}{\partial \theta^3} + |a|^2a \\
                                        &+ F_{\rm pmp} + F_{\rm ref}e^{i(\alpha_{\rm ref} - \alpha_{\rm pmp} + D_{\rm int}(\mu_{\rm ref}))t + i\mu_{\rm ref}\theta},
    \end{split}
\end{equation}
where $D_3$ is the third-order dispersion coefficient. Normalizing \cref{eq:mlle_D3} with respect to the detuning $\alpha_{\rm pmp}$, we now find
\begin{equation}\label{eq:mlle_perturbed_D3}
    \begin{split}
        \frac{\partial \psi}{\partial t} = -\frac{i}{2}\psi + \frac{i}{2}\frac{\partial^2 \psi}{\partial x^2} + i|\psi|^2\psi + \epsilon\mathcal{P}^{\rm 3OD}(x,t),
    \end{split}
\end{equation}
where $\epsilon\mathcal{P}^{\rm 3OD}(x,t)$ contain the terms we treat perturbatively, given by
\begin{equation}\label{eq:perturbation_D3}
    \epsilon\mathcal{P}^{\rm 3OD}(x,t) = -\delta\psi + \beta_3 \frac{\partial^3 \psi}{\partial x^3} + \sigma\frac{\partial \psi}{\partial x} - \frac{i}{2}h_{\rm pmp} - \frac{i}{2}h_{\rm ref}e^{i\mu_s x},
\end{equation}
where $\beta_3$ is the detuning-normalized third-order dispersion coefficient. We choose the signs of $\beta_3$ and $\mu_s$ such that the DW mode and the reference pump mode are on opposite sides of the spectrum. In this modified system, there are now two routes for the conservation of the soliton spectral center of mass in the KIS regime---a spectral recoil similar to what was described in the case of a pure quadratic integrated dispersion, and also through an enhancement of the DW power. That is, the presence of a DW acts as a new degree of freedom for the soliton self-balancing to occur. 

We first calculate the impact of KIS on the soliton spectral recoil. Assuming that we are still in the perturbed-NLSE limit, we use the soliton ansatz that was derived in \cite{kodama_nonlinear_1987} and used in \cite{akhmediev_cherenkov_1995},
\begin{equation}\label{eq:ansatz_D3}
    \psi_s(x,t) = A\sech[A(x-x_0)]e^{[i\mu_r^{\rm 3OD}(x-x_0)- 3iA\tanh(A(x-x_0))]},
\end{equation}
where we have applied the transformation $x = x - \beta_3t$ to remove the change to the group velocity caused by third-order dispersion, and $\mu_r^{\rm 3OD}$ is the soliton central frequency in the presence of third-order dispersion and KIS. Using the same perturbative technique that we used in the previous section but with the modified perturbation \cref{eq:perturbation_D3}, we obtain an expression for the spectral recoil with third-order dispersion and a reference pump
\begin{equation}\label{eq:recoil_D3}
    h_\mathrm{DKS}^2\mu_r^{\rm 3OD} = \sigma - \frac{h_{\rm ref}}{4}\frac{\pi^2}{A^2}\sech\left( \frac{\mu_s \pi}{2A} \right)\tanh\left( \frac{\mu_s \pi}{2A} \right) + 2\beta_3A^2,
\end{equation}
which is identical to \cref{eq:self_balancing_D2} except for the additional term $2\beta_3A^2$ which is the spectral recoil due to the third-order dispersion term. First, we note that \cref{eq:recoil_D3} predicts a shift in the soliton spectral center of mass with a magnitude proportional to $\beta_3$ which is a result of the change in DKS momentum due to the emission of a DW. \cref{eq:recoil_D3} also shows that the presence of the reference pump continues to modify the soliton spectral center of mass in a manner identical to what was observed in the simplistic pure-$D_2$ system that was considered in the previous section. Additionally, \cref{eq:recoil_D3} shows that injecting the reference pump at a mode on the opposite side of the spectrum to where the DW is can counteract the recoil due to the DW. This effect provides a KIS-enabled all-optical pathway to suppress unwanted spectral recoil, like those arising through the Raman effect or strong DW emissions. We verified the validity of this effect through MLLE simulations and observed that synchronizing the DKS with a reference pump with sufficiently high power and at a mode on the opposite side of the spectrum as the DW can completely cancel the recoil induced by the DW emission, as shown in \cref{fig:3}(a)-(b). The MLLE simulations also show that the power of the comb teeth increase around the DW mode [\cref{fig:3}(c)]. Having numerically and analytically shown that KIS can lead to a spectral recoil, we next calculate the DW power as a function of mode number (see Supplementary Material),
\begin{equation}\label{eq:delta_dw_power}
\begin{split}
    \Delta P(\mu_{\rm DW}) = C(\beta_3,\mu_{\rm DW})\frac{h_{\rm ref}\pi^4}{8A^3}\sech\left(\frac{\pi\mu_s}{2A}\right)\tanh\left(\frac{\pi\mu_s}{2A}\right),
\end{split}
\end{equation}
where ${C(\beta_3,\mu_{\rm DW}) = \beta_3\mu_{\rm DW}^3 \sech({\pi\mu_{\rm DW}}/{2A})\tanh(\pi\mu_{\rm DW}/{2A})}$ and $\mu_{\rm DW}$ is the relative mode number at which the DW is found. \cref{eq:delta_dw_power} implies that there exists an optimal mode number $\mu_{s,\rm opt}$ for injecting the reference pump to achieve maximal DW power enhancement which is independent of the third-order dispersion coefficient $\beta_3$, and therefore independent of the DW mode number. We next perform numerical simulations where we fix the reference pump amplitude and calculate the change in DW power for a range of $\mu_s$, and repeat this for different reference pump amplitudes, shown in \cref{fig:3}(d). By fitting the numerically obtained results with the analytical expression in \cref{eq:delta_dw_power}, we find that the DW power follows the predicted $\sech(\cdot)\tanh(\cdot)$ trend accurately. In agreement with \cref{eq:delta_dw_power}, we observe that the mode number with optimal DW enhancement does not change with the reference pump amplitude. As $\mu_{s,\rm opt} \neq -\mu_{\rm DW}$, \cref{eq:delta_dw_power} demonstrates that the DW enhancement is not a result of simple four-wave mixing, but is instead due to the soliton conserving its spectral center of mass via a combination of spectral recoil and DW power enhancement. Next, we calculate the change in power of the comb tooth at the mode pumped by the reference pump (see Supplementrary Material),
\begin{equation}\label{eq:delta_pref}
    \Delta P(\mu_s) = \beta_3 \mu_s^3 \frac{h_{\rm ref}\pi^4}{8A^3}\sech^2\left( \frac{\mu_s \pi}{2A}\right)\tanh^2\left( \frac{\mu_s \pi}{2A}\right).
\end{equation}
Dividing \cref{eq:delta_dw_power} by \cref{eq:delta_pref}, we obtain
\begin{equation}\label{eq:ref_to_dw_efficiency}
    \frac{\Delta P(\mu_{\rm DW})}{\Delta P(\mu_{s})} = \frac{\mu_{\rm DW}^3 \sech\left( \frac{\mu_{\rm DW} \pi}{2A}\right)\tanh\left( \frac{\mu_{\rm DW} \pi}{2A}\right)}{\mu_s^3 \sech\left( \frac{\mu_s \pi}{2A}\right)\tanh\left( \frac{\mu_s \pi}{2A}\right)},
\end{equation}
which equals the reference-to-DW conversion efficiency via self-balancing. Rearranging the terms in \cref{eq:delta_dw_power}, we obtain
\begin{equation}\begin{split}\label{eq:self_balancing_D3}
    &\frac{1}{\mu_{\rm DW}^3}\cosh\left( \frac{\mu_{\rm DW} \pi}{2A}\right)\coth\left( \frac{\mu_{\rm DW} \pi}{2A}\right)\Delta P(\mu_{\rm DW}) \\ &\hspace{1cm}= \frac{1}{\mu_{\rm s}^3}{\cosh\left( \frac{\mu_{s} \pi}{2A}\right)\coth\left( \frac{\mu_s \pi}{2A}\right)}\Delta P(\mu_{s}). 
\end{split}\end{equation}
Noting that $(1/x^3)\cosh(x)\coth(x)$ is always a positive quantity and increases monotonically at large $\vert x \vert$, we can draw an analogy between \cref{eq:self_balancing_D3} and the law of moments and re-write \cref{eq:self_balancing_D3} as
\begin{equation}\label{eq:law_of_moments_analogy}
    m(\mu_{\rm DW})l(\mu_{\rm DW}) = m(\mu_s)l(\mu_s),
\end{equation}
where ${m(\mu_n) = \Delta P(\mu_n)}$ represents a mass-like quantity and ${l(\mu_n)=(1/\mu_n^3){\cosh({\mu_{n} \pi}/{2A})\coth({\mu_n \pi}/{2A})}}$ represents a length-like quantity. To verify the validity of \cref{eq:law_of_moments_analogy}, we simulate the MLLE (\cref{eq:mlle_D3}) with a fixed reference pump power and calculate the ratio in \cref{eq:ref_to_dw_efficiency} for increasing $\mu_s$. We observe that the ratio becomes larger at high $\mu_s$, as shown in \cref{fig:3}(e) and predicted by \cref{eq:law_of_moments_analogy}. 
%   ___            
%  | __|__ __ _ __ 
%  | _| \ \ /| '_ \
%  |___|/_\_\| .__/
%            |_|   
\section{Numerical and experimental demonstration of dispersive wave enhancement}

Having established the self-balancing behavior in purely quadratic and in cubic integrated dispersion systems, we now extend our analysis to a resonator with an experimentally realistic and desirable quartic dispersion that supports two distinct DWs on either side of the main pump frequency. This system provides a practical platform for achieving broadband, octave-spanning frequency combs suitable for self-referencing applications, where high power at both spectral extremes is desirable to realize sufficient signal-to-noise ratio in $f_{\rm ceo}$ detection. While dispersion engineering can introduce DWs at either end of the soliton spectrum, their powers, at higher frequencies, are limited due to inefficient out-coupling. The self-balancing mechanism that KIS enables offers a means to redistribute soliton energy, enabling selective enhancement of the weaker high-frequency DW without perturbing the soliton repetition rate. We model the behavior by employing the MLLE (\cref{eq:mlle}) with an integrated dispersion up to fourth-order terms, thereby generating a microcomb that exhibits two DWs unequally spaced from the primary pump frequency, as shown in \cref{fig:4}(a). When the reference pump is injected at a frequency that coincides with the frequency of the low-frequency DW, the soliton enters the KIS regime, fixing its repetition rate while enforcing the conservation of its spectral center of mass. Due to self-balancing, the soliton re-allocates its energy by enhancing the high-frequency DW power, as shown in \cref{fig:4}(a). Next, we consier how the high-frequency DW power depends on the reference pump power. \cref{eq:delta_dw_power} predicts that the change in DW power is linearly proportional to the change in the power of the mode pumped by the reference pump. In the MLLE, we calculate the change in the high-frequency DW power as a function of change in the reference pump power, shown in \cref{fig:4}(b). We observe that the DW power follows the linear relationship that is predicted by the soliton perturbation theory model.

Experimentally, we generate an octave-spanning DKS microcomb by using a \qty{670}{\nm} thick, fully SiO$_2$-clad \ce{Si3N4} microring resonator fabricated in a foundry and with a ring width of \qty{RW=860}{\nm} similar to\ that in ref.~\cite{MoilleOptica2025}.
We use a \qty{284.3}{\THz} pump laser at \qty{125}{\mW} on-chip power to generate the DKS, where the cavity is thermally stabilized by a counter-propagating \qty{309}{\THz} cooler for adiabatic access to the DKS~\cite{ZhouLightSciAppl2019,ZhangOptica2019}. %
The resonator dispersion enables octave-spanning operation with dispersive wave generation at \qty{190.4}{\THz} and \qty{388.3}{\THz}~[\cref{fig:4}(c)]. %
We use a reference laser at \qty{191.4}{\THz}, offset by one comb tooth from the low-frequency dispersive wave for on-resonance operation and DKS optimization. %
With \qty{10}{\mW} on-chip reference power, the higher frequency DW increases by \qty{22}{\dB}, yielding a comb tooth at \qty{389.3}{\THz} exceeding \qty{2.5}{\uW} (\cref{fig:4}c). %
We note that the maximum comb tooth of the short DW in the KIS case is offset by one compared to the unperturbed DKS, likely due to imperfect reference pump centering in the KIS window and a slight repetition rate change that alters phase matching with the cavity resonances. %
Varying the reference on-chip power from about \qty{0.2}{\mW} to \qty{9}{\mW} and filtering only the comb tooth at \qty{389.3}{\THz}, we observe a trend similar to the theoretical prediction, despite variations in input coupling (\cref{fig:4}d). %

%    ___                 _ 
%   / __| ___  _ _   __ | |
%  | (__ / _ \| ' \ / _|| |
%   \___|\___/|_||_|\__||_|
\section{Conclusion}
We have theoretically and experimentally demonstrated KIS-induced self-balancing of dissipative Kerr solitons. Through perturbative analysis of the multi-pump Lugiato-Lefever equation and corresponding numerical simulations, we showed that a soliton can conserve its spectral center of mass by dynamically redistributing energy between different regions of its spectrum. In the simplest quadratic-dispersion system, this manifests as a compensating spectral recoil on the opposite side of the optical spectrum to the reference pump. When higher-order dispersion is included, this same self-balancing mechanism transfers energy into a DW on the opposite side of the spectrum, leading to measurable power enhancement without changing the repetition rate. Extending the analysis to a quartic-dispersion resonator supporting two DWs, we showed that synchronizing the soliton at the low-frequency DW results in a self-balancing-induced enhancement of the high-frequency DW. This redistribution, verified numerically and confirmed experimentally, yields up to a \qty{22}{\dB} increase in high-frequency DW power. The observed linear scaling of DW power with reference pump power further demonstrates the presence of self-balancing  and distinguishes the process from simple four-wave mixing. These results establish soliton self-balancing as a universal mechanism governing the energy flow in DKSs operating in the KIS regime. Beyond revealing a new form of nonlinear cavity equilibrium and intracavity energy dynamics, self-balancing enables the formation of optical spectra with high-power comb teeth at the spectral extrema, which provides a pathway to effectively detect the CEO frequency with a high signal-to-noise ratio through {$f$-$2f$} self-referencing. 

\subsection*{Acknowledgments}
The photonics chips were fabricated following the process presented in~\cite{MoilleNature2023}. P.S. and C.M. acknowledge support from National Center for Manufacturing Sciences (Cooperative Agreements 2022138-142232 and 2023200-142386 as sub-awards from DoD Cooperative Agreements HQ0034-20-2-0007 and HQ0034-24-2-0001) and from the National Institute of Standards and Technology (Grant No. 60NANB24D106). %
G.M. and K.S. acknowledge partial funding support from the Space Vehicles Directorate of the Air Force Research Laboratory and the NIST-on-a-chip program of the National Institute of Standards and Technology. %
We thank Shao-Chien Ou and Lida Xu for insightful feedback. 

\subsection*{Author contributions}
P.S. developed the theory and performed the simulations. G.M. led the project and performed the experiments. %
C.M. and K.S. contributed in the understanding of the physical phenomenon. %
All the authors wrote the manuscript. %
All the authors contributed and discussed the content of this manuscript.

\subsection*{Competing interests}
G.M., C.M., and K.S have submitted a provisional patent application based on aspects of the work presented in this paper.

\subsection*{Additional Information}
Correspondence and requests for materials should be addressed to G.M.

\subsection*{Data availability}
The data that supports the plots within this paper and other findings of this study are available from the corresponding authors upon reasonable request.

% \clearpage
%
\clearpage
\input{StandaloneSupMat}

\end{document}

%% file: StandaloneSupMat.tex
\onecolumngrid
\appendix
\renewcommand{\thesection}{\arabic{section}}
\titleformat{\section}{\large\bfseries}{S.\thesection}{10pt}{}[]
\renewcommand\thefigure{S.\arabic{figure}}  
\renewcommand\thetable{S.\Roman{table}}
\renewcommand{\arraystretch}{1.3}
\setcounter{figure}{0} 
\setcounter{equation}{0} 
% \numberwithin{equation}{section}
\renewcommand{\theequation}{S.\arabic{section}.\arabic{equation}}
\renewcommand{\appendixpagename}{%
\centering%
\Large
Supplementary Information: \mytitle
}
\appendixpage

\section{\label{supsec:review_soliton_perturbation_theory}Review of soliton perturbation theory}
\medskip
In this section, we will review how soliton perturbation theory can be used to calculate the impact of perturbations on soliton parameters. The nonlinear Schr\"odinger equation (NLSE) can be written as
\begin{equation}\label{supeq:NLSE}
    \frac{\partial \psi}{\partial t} \equiv N(\psi) = -\frac{i}{2}(A^2 + \mu_r^2)\psi + \mu_r \frac{\partial \psi}{\partial x} + \frac{i}{2}\frac{\partial^2 \psi}{\partial x^2} + i|\psi|^2\psi,
\end{equation}
which admits stationary soliton solutions of the form
\begin{equation}\label{supeq:SS}
    \psi_s(x,t) = A\sech[A(x-x_0)]e^{i[\mu_r (x-x_0)] - i\phi},
\end{equation}
where the parameters $p_j = \{A, x_0, \mu_r, \textrm{and } \phi\}$ are the soliton amplitude, central position, central frequency (or spectral recoil), and phase respectively. The linearized operator of the NLSE, ${\mathcal{L} = \left.\delta N/\delta \Psi\right|_{\Psi_s}}$ is 
\begin{equation}\label{supeq:linearizedNLSE}
    \mathcal{L} = \begin{pmatrix}
        -\frac{i}{2}(A^2 + \mu_r^2) + \mu_r \frac{\partial}{\partial x} + \frac{i}{2}\frac{\partial^2}{\partial x^2} + 2i|\psi_s|^2 & i\psi_s^2 \\
        -i(\psi_s^*)^2 & \frac{i}{2}(A^2 + \mu_r^2) + \mu_r \frac{\partial}{\partial x} - \frac{i}{2}\frac{\partial^2}{\partial x^2} - 2i|\psi_s|^2 
    \end{pmatrix}
\end{equation}
and $\Psi = (\psi,\psi^*)$. Since the soliton solution in \cref{supeq:SS} has four free parameters, \cref{supeq:linearizedNLSE} has four generalized eigenfunctions with zero eigenvalues,
\begin{subequations}\label{supeq:linearized_eigenfunctions}
    \begin{align}
        \psi_A &= \frac{\psi_s}{A}\{1 - A(x-x_0)\tanh[A(x-x_0)]\}\\
        \psi_{\mu_r} &= i(x-x_0)\psi_s\\
        \psi_\phi &= -i\psi_s\\
        \psi_{x_0} &= -\frac{\partial \psi_s}{\partial x} = \psi_s\{-i\mu_r + A\tanh[A(x-x_0)]\}
    \end{align}
\end{subequations}
each of which corresponds to a perturbation of one of the four soliton parameters. To find the adjoint eigenfunctions of \cref{supeq:linearized_eigenfunctions}, we will first write the adjoint of \cref{supeq:linearizedNLSE}. The adjoint matrix $\mathcal{L}^\dagger$ is given by
\begin{equation}\label{supeq:adjoint}
    \mathcal{L}^\dagger = \begin{pmatrix}
        \frac{i}{2}(A^2 + \mu_r^2) + \mu_r \frac{\partial}{\partial x} - \frac{i}{2}\frac{\partial^2}{\partial x^2} - 2i|\psi_s|^2 & i\psi_s^2 \\
        -i(\psi_s^*)^2  &  -\frac{i}{2}(A^2 + \mu_r^2) + \mu_r \frac{\partial}{\partial x} + \frac{i}{2}\frac{\partial^2}{\partial x^2} + 2i|\psi_s|^2 
    \end{pmatrix}.
\end{equation}
The generalized eigenfunctions of \cref{supeq:adjoint} that satisfy the inner product relation $\left< \overline{\Psi}_m,\Psi_n\right> = \delta_{mn}$ for the eigenfunctions of \cref{supeq:linearizedNLSE} are
\begin{subequations}\label{supeq:adjoint_eigenfunctions}
    \begin{align}
        \overline{\psi}_A &= \psi_s\\
        \overline{\psi}_{\mu_r} &= i\psi_s\tanh[A(x-x_0)]\\
        \overline{\psi}_\phi &= -\frac{\psi_s}{A}\{i + \mu_r(x-x_0) - iA(x-x_0)\tanh[A(x-x_0)]\}\\
        \overline{\psi}_{x_0} &= (x-x_0)\frac{\psi_s}{A}.
    \end{align}
\end{subequations}
Next, we assume that the Lugiato-Lefever equation (LLE) is of the form
\begin{equation}\label{supeq:normalizedLLE}
    \frac{\partial \psi}{\partial t} = -\frac{i}{2}\psi + \frac{i}{2}\frac{\partial^2 \psi}{\partial x^2} + i|\psi|^2\psi + \epsilon\chi(x,t,\psi)
\end{equation}
where $\chi(x,t,\psi)$ contains all loss and gain (pump) terms, and any other perturbation whose impact we want to analyze that cannot otherwise be handled analytically. We can rewrite \cref{supeq:normalizedLLE} in the form of \cref{supeq:NLSE} as
\begin{equation}\label{supeq:NLSE2LLE}
   \frac{\partial \psi}{\partial t} = N(\psi) - \mu_r\frac{\partial \psi}{\partial x} + \frac{i}{2}(A^2 + \mu_r^2 - 1)\psi + \epsilon\chi(x,t,\psi),
\end{equation}
and equivalently in matrix form as
\begin{equation}\label{supeq:NLSE2LLE_matrix}
   \frac{\partial \Psi}{\partial t} = N(\Psi) - \mu_r\frac{\partial \Psi}{\partial x} + \frac{i}{2}(A^2 + \mu_r^2 - 1)I\Psi + \epsilon \mathcal{P}(x,t,\Psi),
\end{equation}
where $I$ is the identity matrix. In general, we can write the solution of the LLE as a sum of the NLSE solution in \cref{supeq:SS} and a perturbative correction term
\begin{equation}
    \Psi(x) = \Psi_s(p_j) + \epsilon\Delta\Psi(x,t),
\end{equation}
Substituting this in \cref{supeq:NLSE2LLE_matrix}, for any $p_j$, we obtain
\begin{align}\label{supeq:GeneralPerturbedEquation}
    \Psi_{p_j}\frac{d p_j}{d t} + \epsilon \frac{\partial \epsilon\Delta\Psi}{\partial t} = &\epsilon\mathcal{L}\Delta\Psi + \mu_r\Psi_{x_0} - \mu_r\frac{\partial \Delta\Psi}{\partial x} - \frac{1}{2}(A^2 + \mu_r^2 - 1)I\Psi_{\phi} \\
    &+ \epsilon\frac{i}{2}(A^2 + \mu_r^2 - 1)I\Delta\Psi + \epsilon\mathcal{P}.\nonumber
\end{align}
In \cref{supeq:GeneralPerturbedEquation}, we have made use of the fact that ${\partial \psi_s}/{\partial x} = -\psi_{x_0}$ and $i\psi_s = -\psi_{\phi}$ to simplify the form of the equation. To obtain the equation of motion of any of the free parameters, we simply have to take the inner product of \cref{supeq:GeneralPerturbedEquation} with respect to the corresponding adjoint eigenfunction and note that ${\left<\overline{\Psi}_m,\Delta\Psi\right> = 0}$ and ${\left<\overline{\Psi}_m,\Psi_n\right> = \delta_{m,n}}$. Substituting $x_0$ and $\mu_r$ in \cref{supeq:GeneralPerturbedEquation}, we obtain the equations of motion for the central position and central frequency,
\begin{subequations}\label{supeq:recoil_and_KIS}
    \begin{align}
        \frac{d x_0}{d t} &= \mu_r + \epsilon \left<\overline{\Psi}_{x_0},\mathcal{P}\right>\\
        \frac{d \mu_r}{d t} &= \epsilon \left<\overline{\Psi}_{\mu_r},\mathcal{P}\right>
    \end{align}
\end{subequations}

\section{\label{supsec:D2_selfBalancing_derivation}Derivation of analytical expression for spectral recoil in pure-$D_2$ microresonator}
\medskip
Here, we derive the self-balancing equation for a purely quadratic integrated dispersion microresonator, given in~Eq.~(7) in the main text. The soliton dynamics in a dual-pumped microresonator can be modeled using the loss-normalized multi-pump Lugiato-Lefever equation (MLLE)~\cite{TaheriEur.Phys.J.D2017,MoilleOptica2025}
\begin{equation}\label{supeq:mlle}
    \begin{split}
        \frac{\partial a}{\partial t} = &-\left( 1 + i\alpha_{\rm pmp} \right)a + i\sum_\mu D_{\rm int}(\mu)A(\mu) e^{i\mu\theta} + i|a|^2a \\
                                        &+ F_{\rm pmp} + F_{\rm ref}e^{i(\alpha_{\rm ref} - \alpha_{\rm pmp} + D_{\rm int}(\mu_{\rm ref}))t + i\mu_{\rm ref}\theta},
    \end{split}
\end{equation}
where $a$ is the intracavity field amplitude, $A$ is its Fourier transform, $t$ is time, $\theta$ is the azimuthal coordinate, $\alpha_{\rm pmp}$ ($\alpha_{\rm ref}$) is the detuning of the primary (reference) pump with corresponding amplitudes $F_{\rm pmp}$ ($F_{\rm ref}$), $D_{\rm int}$ is the integrated dispersion as a function of the relative mode number $\mu$, and $\mu_s$ is the relative mode number of the reference pump. Considering only up to second-order dispersion, \cref{supeq:mlle} reduces to 
\begin{equation}\label{supeq:mlle_D2}
    \begin{split}
                \frac{\partial a}{\partial t} = &-\left( 1 + i\alpha_{\rm pmp} \right)a + iD_2 \frac{\partial^2 a}{\partial \theta^2} + i|a|^2a \\
                                        &+ F_{\rm pmp} + F_{\rm ref}e^{i(\alpha_{\rm ref} - \alpha_{\rm pmp} + D_{\rm int}(\mu_{\rm ref}))t + i\mu_{\rm ref}\theta},
    \end{split}
\end{equation}
where $D_2$ is the second-order dispersion coefficient. The loss-normalized MLLE is convenient for physical intuition and numerical simulations because it directly relates to experimentally tunable parameters, namely the pump amplitudes ($F_{\rm pmp}$ and $F_{\rm ref}$) and pump detunings ($\alpha_{\rm pmp}$ and $\alpha_{\rm ref}$). However, \cref{supeq:mlle} cannot be directly used to analytically study the impact of the reference pump on the soliton spectral energy dynamics. To facilitate analytical treatment of the energy dynamics, we normalize \cref{supeq:mlle_D2} with respect to the main pump detuning and obtain~\cite{Mizrahi2024}
\begin{equation}\label{supeq:mlle_det_normalized}
    \begin{split}
        \frac{\partial \psi}{\partial t} = &-\left(\delta + \frac{i}{2} \right)\psi + \frac{i}{2}\frac{\partial^2 \psi}{\partial x^2} + i|\psi|^2\psi \\
        &- \frac{i}{2}h_{\rm pmp} - \frac{i}{2}h_{\rm ref}e^{i\mu_s x} - \sigma \frac{\partial \psi}{\partial x},
    \end{split}
\end{equation}
where ${\psi=a/\sqrt{2\alpha_{\rm pmp}}}$ is the normalized intracavity field amplitude, ${\delta=1/2\alpha_{\rm pmp}}$ is the normalized loss, ${h_{\rm {pmp,ref}} = iF_{\rm {pmp,ref}}/\sqrt{2}\alpha_{\rm pmp}^{3/2}}$ is the normalized main (reference) pump amplitude, ${t=t\times2\alpha_{\rm pmp}}$ is the normalized time, ${x = \sqrt{2\alpha_{\rm pmp}}\theta}$, and $\sigma$ is the normalized detuning of the reference pump. In the large main pump detuning limit ($\alpha_{\rm pmp}\rightarrow\infty$), the normalized loss ($\delta$), and gain terms ($h_{\rm pmp}$ and $h_{\rm ref}$) become small, and can be treated as perturbations to the NLSE. Therefore, we can re-write \cref{supeq:mlle_det_normalized} as
\begin{equation}\label{supeq:rearrLLE}
    \frac{\partial \psi}{\partial t} = -\frac{i}{2}\psi + \frac{i}{2}\frac{\partial^2 \psi}{\partial x^2} + i|\psi|^2\psi + \epsilon\chi,
\end{equation}
where $\epsilon\chi = -\delta\psi   + \sigma{\partial \psi}/{\partial x} - ({i}/{2})(h_\mathrm{main} + h_\mathrm{ref}e^{i\mu_s x})$, and the coefficient $\epsilon$ is used to denote that the terms in $\epsilon\chi$ are small. \cref{supeq:rearrLLE} is exactly the same as \cref{supeq:normalizedLLE}, and therefore we can use \cref{supeq:recoil_and_KIS} to obtain the equations of motion. For the central position, we obtain
\begin{align}\label{supeq:position_ODE_exact}
    \frac{\partial x_0}{\partial t} &= \frac{\pi}{A}M_S\mu_r - \sigma - \frac{h_{\mathrm{ref}}}{2}\sin(\mu_s x_0)\int_{-\infty}^{\infty} x \sech(Ax)\sin((\mu_s + \mu_r)x)dx  \\
                                    &= \frac{\pi}{A}M_s\mu_r - \sigma - \frac{h_{\mathrm{ref}}\pi^2}{4A^2}\sech\left[\frac{(\mu_s + \mu_r) \pi}{2A}\right]\tanh\left[\frac{(\mu_s + \mu_r) \pi}{2A}\right] \nonumber
\end{align}
Since the spectral recoil is typically much smaller than the reference pump mode number, we assume that ${\mu_s\gg \mu_r}$ and simplify \cref{supeq:position_ODE_exact} as
\begin{align}\label{supeq:position_ODE}
    \frac{\partial x_0}{\partial t} &= \frac{\pi}{A}M_s\mu_r - \sigma - \frac{h_{\mathrm{ref}}\pi^2}{4A^2}\sech\left(\frac{\mu_s \pi}{2A}\right)\tanh\left(\frac{\mu_s\pi}{2A}\right).
\end{align}
The coefficient $({\pi}/{A})M_S$ is equal to one and ${M_S = \int_{-\infty}^{\infty}|\mathcal{F}(\psi(x))|^2d\mu}$ is the soliton's spectral mass. In the KIS regime, the soliton solution is stationary. Therefore, setting ${{\partial x_0}/{\partial t} = 0}$, we obtain
\begin{equation}\label{supeq:self_balancing_D2_sup}
    \frac{\pi}{A}M_S\mu_r = \frac{h_{\mathrm{ref}}\pi^2}{4A^2}\sech\left(\frac{\mu_s \pi}{2A}\right)\tanh\left(\frac{\mu_s \pi}{2A}\right) + \sigma,
\end{equation}
which is~Eq.~(7) in the main text. 
\medskip
\section{\label{supsec:D3_selfBalancing_derivation}Derivation of analytical expression for dispersive wave enhancement in the presence of third-order dispersion}
\medskip
In this section, we derive the enhancement in the dispersive wave (DW) power due to self-balancing (Eq.~(13) in the main text). The MLLE in the presence of third-order dispersion is
\begin{equation}\label{supeq:mlle_perturbed_D3}
    \begin{split}
        \frac{\partial \psi}{\partial t} = -\frac{i}{2}\psi + \frac{i}{2}\frac{\partial^2 \psi}{\partial x^2} + i|\psi|^2\psi + \epsilon\mathcal{P}^{\rm 3OD}(x,t),
    \end{split}
\end{equation}
where $\epsilon\mathcal{P}^{\rm 3OD}(x,t)$ contain the terms we treat perturbatively, given by
\begin{equation}\label{supeq:perturbation_D3}
    \epsilon\mathcal{P}^{\rm 3OD}(x,t) = -\delta\psi + \beta_3 \frac{\partial^3 \psi}{\partial x^3} + \sigma\frac{\partial \psi}{\partial x} - \frac{i}{2}h_{\rm pmp} - \frac{i}{2}h_{\rm ref}e^{i\mu_s x},
\end{equation}
where $\beta_3$ is the detuning-normalized third-order dispersion coefficient. To use soliton perturbation theory, we use the expression derived in \cite{akhmediev_cherenkov_1995} as the soliton solution ansatz in the presence of third-order dispersion
\begin{equation}\label{supeq:SS_TOD}
        \psi_s^{\mathrm{3OD}}(x,t) = A\sech [A(x - \beta_3 t)]e^{i\beta_3 \{2A^2(x - \beta_3 t) - 3A\tanh[A(x - \beta_3 t)]\}}.
\end{equation}
First, we calculate the impact of KIS on the spectral recoil $\mu_r$ in \cref{supeq:mlle_perturbed_D3}. Using the ansatz in \cref{supeq:SS_TOD} and subtituting \cref{supeq:perturbation_D3} in \cref{supeq:recoil_and_KIS}, and setting ${dx_0/dt=0}$, we obtain
\begin{equation}\label{supeq:recoil_TOD}
    \mu_r^{\mathrm{3OD}} = \sigma + \frac{h_{\mathrm{ref}}}{4}\frac{\pi^2}{A^2}\sech(\frac{\mu_s \pi}{2A})\tanh(\frac{\mu_s \pi}{2A}) + 2\beta_3 A^2,
\end{equation}
where the term $2\beta_3A^2$ is the spectral recoil due to third-order dispersion and the remaining terms are due to the self-balancing effect. Having obtained the spectral recoil with KIS in the presence of third-order dispersion, we modify the ansatz in \cref{supeq:SS_TOD} to include $\mu_r^{(\mathrm{3OD})}$ explicitly,
\begin{align}\label{supeq:SS_TOD_Modified}
    \psi_s^{\textrm{3OD}} &= A\sech(Ax')\exp[i\mu_r^{(\mathrm{3OD})}x' - 3A\epsilon\beta_3\tanh(Ax)] \\
    &\approx A\sech(Ax)[1 + i\mu_r^{\textrm{3OD}}x - 3iA\epsilon\beta_3\tanh(Ax)].\nonumber
\end{align}
where $x' = x - x_0$. We will now use the modified ansatz in \cref{supeq:SS_TOD_Modified} to calculate the position and power of the DW in the presence of third-order dispersion following \cite{akhmediev_cherenkov_1995}. DWs occur when the soliton is in resonance with a linear wave, i.e., when the soliton wavenumber is equal to that of a linear dispersive wave. Assuming that the DW is a linear wave of the form $e^{i(k_lt + \mu x)}$ and substituting it into \cref{supeq:NLSE}, we obtain
\begin{equation}\label{supeq:klin}
    k_l(\mu) = -\frac{1}{2}\mu^2 - \beta_3 \mu^3.
\end{equation}
Equating the wavenumber obtained in \cref{supeq:klin} to the soliton wavenumber ($k_s = A^2 / 2$), we obtain the DW phase-matching condition
\begin{equation}\label{supeq:phase_match}
    \beta_3 \mu_{\mathrm{DW}}^3 + \frac{1}{2}\mu_{\mathrm{DW}}^2 + \frac{A^2}{2} = 0. 
\end{equation}
Solving \cref{supeq:phase_match} to first order, we obtain the DW position
\begin{equation}\label{supeq:DW_pos}
    \mu_{\mathrm{DW}} = -\frac{1}{2\beta_3}.
\end{equation}
We assume that the solution of \cref{supeq:mlle_perturbed_D3} is of the form $\psi(x,t) = \psi_s^{\mathrm{3OD}}(x,t) + f(x,t)$, where $f(x,t)$ is the DW. Linearizing \cref{supeq:mlle_perturbed_D3} with respect to $f(x,t)$, we obtain
\begin{equation}\label{supeq:linearized_TOD}
    \frac{\partial f}{\partial t} - \frac{i}{2}\frac{\partial^2 f}{\partial x^2} + i\beta_3 \frac{\partial^3 f}{\partial x^3} - {2i|\psi_s^{\mathrm{3OD}}|^2 f -  i(\psi_s^{\mathrm{3OD}})^2 f} = -i\beta_3\frac{\partial^3 \psi_s^{\mathrm{3OD}}}{\partial x^3}.
\end{equation}
Assuming that the DW has a fixed azimuthal form that is phase-matched with the soliton, we can substitute the ansatz $f(x,t) = r(x)e^{ik_s t}$ in \cref{supeq:linearized_TOD} and take the Fourier transform to obtain
\begin{equation}\label{supeq:DW_Fourier}
    R(\mu) = \frac{\mathcal{F}[\beta_3 \partial^3 \psi_s^{\mathrm{3OD}} / \partial x^3]}{k_l(\mu) - k_s} \equiv \frac{P(\mu)}{k_l(\mu) - k_s},
\end{equation}
where $R(\mu)$ is the DW in the normalized mode number domain, $\mathcal{F}$ is the Fourier transform operator, and $P(\mu)$, which is proportional to the DW power, is
\begin{align}\label{supeq:power_mu}
    P(\mu,h_{\rm ref}) = -i\beta_3\mu^3 \left[ \pi\sech(\frac{\pi\mu}{2A}) + \mu^{\textrm{3OD}}(\beta_3,h_{\rm ref})\frac{\pi^2}{2A}\sech(\frac{\pi\mu}{2A})\tanh(\frac{\pi\mu}{2A}) - 3\beta_3\pi\mu\sech(\frac{\pi\mu}{2A}) \right].
\end{align}
Subsituting \cref{supeq:recoil_TOD} with and without $h_{\rm ref}$ in \cref{supeq:power_mu}, we obtain the DW power enhancement in KIS
\begin{equation}\label{supeq:delta_dw_power}
\begin{split}
    \Delta P(\mu_{\rm DW}) = C(\beta_3,\mu_{\rm DW})\frac{h_{\rm ref}\pi^4}{8A^3}\sech\left(\frac{\pi\mu_s}{2A}\right)\tanh\left(\frac{\pi\mu_s}{2A}\right),
\end{split}
\end{equation}
where $C(\beta_3,\mu_{\rm DW}) = \beta_3\mu_{\rm DW}^3 \sech\left[({\pi\mu_{\rm DW}})/({2A})\right]\tanh\left[({\pi\mu_{\rm DW}})/({2A})\right]$.

%% file: 2024-PAPER-KIS_selfBalacing.bbl
\begin{thebibliography}{30}%
\makeatletter
\providecommand \@ifxundefined [1]{%
 \@ifx{#1\undefined}
}%
\providecommand \@ifnum [1]{%
 \ifnum #1\expandafter \@firstoftwo
 \else \expandafter \@secondoftwo
 \fi
}%
\providecommand \@ifx [1]{%
 \ifx #1\expandafter \@firstoftwo
 \else \expandafter \@secondoftwo
 \fi
}%
\providecommand \natexlab [1]{#1}%
\providecommand \enquote  [1]{``#1''}%
\providecommand \bibnamefont  [1]{#1}%
\providecommand \bibfnamefont [1]{#1}%
\providecommand \citenamefont [1]{#1}%
\providecommand \href@noop [0]{\@secondoftwo}%
\providecommand \href [0]{\begingroup \@sanitize@url \@href}%
\providecommand \@href[1]{\@@startlink{#1}\@@href}%
\providecommand \@@href[1]{\endgroup#1\@@endlink}%
\providecommand \@sanitize@url [0]{\catcode `\\12\catcode `\$12\catcode
  `\&12\catcode `\#12\catcode `\^12\catcode `\_12\catcode `\%12\relax}%
\providecommand \@@startlink[1]{}%
\providecommand \@@endlink[0]{}%
\providecommand \url  [0]{\begingroup\@sanitize@url \@url }%
\providecommand \@url [1]{\endgroup\@href {#1}{\urlprefix }}%
\providecommand \urlprefix  [0]{URL }%
\providecommand \Eprint [0]{\href }%
\providecommand \doibase [0]{https://doi.org/}%
\providecommand \selectlanguage [0]{\@gobble}%
\providecommand \bibinfo  [0]{\@secondoftwo}%
\providecommand \bibfield  [0]{\@secondoftwo}%
\providecommand \translation [1]{[#1]}%
\providecommand \BibitemOpen [0]{}%
\providecommand \bibitemStop [0]{}%
\providecommand \bibitemNoStop [0]{.\EOS\space}%
\providecommand \EOS [0]{\spacefactor3000\relax}%
\providecommand \BibitemShut  [1]{\csname bibitem#1\endcsname}%
\let\auto@bib@innerbib\@empty
%</preamble>
\bibitem [{\citenamefont {Stern}\ \emph {et~al.}(2018)\citenamefont {Stern},
  \citenamefont {Ji}, \citenamefont {Okawachi}, \citenamefont {Gaeta},\ and\
  \citenamefont {Lipson}}]{SternNature2018}%
  \BibitemOpen
  \bibfield  {author} {\bibinfo {author} {\bibfnamefont {B.}~\bibnamefont
  {Stern}}, \bibinfo {author} {\bibfnamefont {X.}~\bibnamefont {Ji}}, \bibinfo
  {author} {\bibfnamefont {Y.}~\bibnamefont {Okawachi}}, \bibinfo {author}
  {\bibfnamefont {A.~L.}\ \bibnamefont {Gaeta}},\ and\ \bibinfo {author}
  {\bibfnamefont {M.}~\bibnamefont {Lipson}},\ }\bibfield  {title} {\bibinfo
  {title} {Battery-operated integrated frequency comb generator},\ }\href
  {https://doi.org/10.1038/s41586-018-0598-9} {\bibfield  {journal} {\bibinfo
  {journal} {Nature}\ }\textbf {\bibinfo {volume} {562}},\ \bibinfo {pages}
  {401} (\bibinfo {year} {2018})}\BibitemShut {NoStop}%
\bibitem [{\citenamefont {Liu}\ \emph {et~al.}(2021)\citenamefont {Liu},
  \citenamefont {Huang}, \citenamefont {Wang}, \citenamefont {He},
  \citenamefont {Raja}, \citenamefont {Liu}, \citenamefont {Engelsen},\ and\
  \citenamefont {Kippenberg}}]{LiuNatCommun2021}%
  \BibitemOpen
  \bibfield  {author} {\bibinfo {author} {\bibfnamefont {J.}~\bibnamefont
  {Liu}}, \bibinfo {author} {\bibfnamefont {G.}~\bibnamefont {Huang}}, \bibinfo
  {author} {\bibfnamefont {R.~N.}\ \bibnamefont {Wang}}, \bibinfo {author}
  {\bibfnamefont {J.}~\bibnamefont {He}}, \bibinfo {author} {\bibfnamefont
  {A.~S.}\ \bibnamefont {Raja}}, \bibinfo {author} {\bibfnamefont
  {T.}~\bibnamefont {Liu}}, \bibinfo {author} {\bibfnamefont {N.~J.}\
  \bibnamefont {Engelsen}},\ and\ \bibinfo {author} {\bibfnamefont {T.~J.}\
  \bibnamefont {Kippenberg}},\ }\bibfield  {title} {\bibinfo {title}
  {High-yield, wafer-scale fabrication of ultralow-loss, dispersion-engineered
  silicon nitride photonic circuits},\ }\href
  {https://doi.org/10.1038/s41467-021-21973-z} {\bibfield  {journal} {\bibinfo
  {journal} {Nature Communications}\ }\textbf {\bibinfo {volume} {12}},\
  \bibinfo {pages} {2236} (\bibinfo {year} {2021})}\BibitemShut {NoStop}%
\bibitem [{\citenamefont {Ou}\ \emph {et~al.}(2025)\citenamefont {Ou},
  \citenamefont {Antohe}, \citenamefont {Carpenter}, \citenamefont {Moille},\
  and\ \citenamefont {Srinivasan}}]{OuOpt.Lett.OL2025}%
  \BibitemOpen
  \bibfield  {author} {\bibinfo {author} {\bibfnamefont {S.-C.}\ \bibnamefont
  {Ou}}, \bibinfo {author} {\bibfnamefont {A.~O.}\ \bibnamefont {Antohe}},
  \bibinfo {author} {\bibfnamefont {L.~G.}\ \bibnamefont {Carpenter}}, \bibinfo
  {author} {\bibfnamefont {G.}~\bibnamefont {Moille}},\ and\ \bibinfo {author}
  {\bibfnamefont {K.}~\bibnamefont {Srinivasan}},\ }\bibfield  {title}
  {\bibinfo {title} {300 mm wafer-scale {{SiN}} platform for broadband soliton
  microcombs compatible with alkali atomic references},\ }\href
  {https://doi.org/10.1364/OL.571893} {\bibfield  {journal} {\bibinfo
  {journal} {Optics Letters}\ }\textbf {\bibinfo {volume} {50}},\ \bibinfo
  {pages} {5578} (\bibinfo {year} {2025})}\BibitemShut {NoStop}%
\bibitem [{\citenamefont {Spencer}\ \emph {et~al.}(2018)\citenamefont
  {Spencer}, \citenamefont {Drake}, \citenamefont {Briles}, \citenamefont
  {Stone}, \citenamefont {Sinclair}, \citenamefont {Fredrick}, \citenamefont
  {Li}, \citenamefont {Westly}, \citenamefont {Ilic}, \citenamefont
  {Bluestone}, \citenamefont {Volet}, \citenamefont {Komljenovic},
  \citenamefont {Chang}, \citenamefont {Lee}, \citenamefont {Oh}, \citenamefont
  {Suh}, \citenamefont {Yang}, \citenamefont {Pfeiffer}, \citenamefont
  {Kippenberg}, \citenamefont {Norberg}, \citenamefont {Theogarajan},
  \citenamefont {Vahala}, \citenamefont {Newbury}, \citenamefont {Srinivasan},
  \citenamefont {Bowers}, \citenamefont {Diddams},\ and\ \citenamefont
  {Papp}}]{SpencerNature2018}%
  \BibitemOpen
  \bibfield  {author} {\bibinfo {author} {\bibfnamefont {D.~T.}\ \bibnamefont
  {Spencer}}, \bibinfo {author} {\bibfnamefont {T.}~\bibnamefont {Drake}},
  \bibinfo {author} {\bibfnamefont {T.~C.}\ \bibnamefont {Briles}}, \bibinfo
  {author} {\bibfnamefont {J.}~\bibnamefont {Stone}}, \bibinfo {author}
  {\bibfnamefont {L.~C.}\ \bibnamefont {Sinclair}}, \bibinfo {author}
  {\bibfnamefont {C.}~\bibnamefont {Fredrick}}, \bibinfo {author}
  {\bibfnamefont {Q.}~\bibnamefont {Li}}, \bibinfo {author} {\bibfnamefont
  {D.}~\bibnamefont {Westly}}, \bibinfo {author} {\bibfnamefont {B.~R.}\
  \bibnamefont {Ilic}}, \bibinfo {author} {\bibfnamefont {A.}~\bibnamefont
  {Bluestone}}, \bibinfo {author} {\bibfnamefont {N.}~\bibnamefont {Volet}},
  \bibinfo {author} {\bibfnamefont {T.}~\bibnamefont {Komljenovic}}, \bibinfo
  {author} {\bibfnamefont {L.}~\bibnamefont {Chang}}, \bibinfo {author}
  {\bibfnamefont {S.~H.}\ \bibnamefont {Lee}}, \bibinfo {author} {\bibfnamefont
  {D.~Y.}\ \bibnamefont {Oh}}, \bibinfo {author} {\bibfnamefont {M.-G.}\
  \bibnamefont {Suh}}, \bibinfo {author} {\bibfnamefont {K.~Y.}\ \bibnamefont
  {Yang}}, \bibinfo {author} {\bibfnamefont {M.~H.~P.}\ \bibnamefont
  {Pfeiffer}}, \bibinfo {author} {\bibfnamefont {T.~J.}\ \bibnamefont
  {Kippenberg}}, \bibinfo {author} {\bibfnamefont {E.}~\bibnamefont {Norberg}},
  \bibinfo {author} {\bibfnamefont {L.}~\bibnamefont {Theogarajan}}, \bibinfo
  {author} {\bibfnamefont {K.}~\bibnamefont {Vahala}}, \bibinfo {author}
  {\bibfnamefont {N.~R.}\ \bibnamefont {Newbury}}, \bibinfo {author}
  {\bibfnamefont {K.}~\bibnamefont {Srinivasan}}, \bibinfo {author}
  {\bibfnamefont {J.~E.}\ \bibnamefont {Bowers}}, \bibinfo {author}
  {\bibfnamefont {S.~A.}\ \bibnamefont {Diddams}},\ and\ \bibinfo {author}
  {\bibfnamefont {S.~B.}\ \bibnamefont {Papp}},\ }\bibfield  {title} {\bibinfo
  {title} {An optical-frequency synthesizer using integrated photonics},\
  }\href {https://doi.org/10.1038/s41586-018-0065-7} {\bibfield  {journal}
  {\bibinfo  {journal} {Nature}\ }\textbf {\bibinfo {volume} {557}},\ \bibinfo
  {pages} {81} (\bibinfo {year} {2018})}\BibitemShut {NoStop}%
\bibitem [{\citenamefont {Newman}\ \emph {et~al.}(2019)\citenamefont {Newman},
  \citenamefont {Maurice}, \citenamefont {Drake}, \citenamefont {Stone},
  \citenamefont {Briles}, \citenamefont {Spencer}, \citenamefont {Fredrick},
  \citenamefont {Li}, \citenamefont {Westly}, \citenamefont {Ilic},
  \citenamefont {Shen}, \citenamefont {Suh}, \citenamefont {Yang},
  \citenamefont {Johnson}, \citenamefont {Johnson}, \citenamefont {Hollberg},
  \citenamefont {Vahala}, \citenamefont {Srinivasan}, \citenamefont {Diddams},
  \citenamefont {Kitching}, \citenamefont {Papp},\ and\ \citenamefont
  {Hummon}}]{NewmanOptica2019}%
  \BibitemOpen
  \bibfield  {author} {\bibinfo {author} {\bibfnamefont {Z.~L.}\ \bibnamefont
  {Newman}}, \bibinfo {author} {\bibfnamefont {V.}~\bibnamefont {Maurice}},
  \bibinfo {author} {\bibfnamefont {T.}~\bibnamefont {Drake}}, \bibinfo
  {author} {\bibfnamefont {J.~R.}\ \bibnamefont {Stone}}, \bibinfo {author}
  {\bibfnamefont {T.~C.}\ \bibnamefont {Briles}}, \bibinfo {author}
  {\bibfnamefont {D.~T.}\ \bibnamefont {Spencer}}, \bibinfo {author}
  {\bibfnamefont {C.}~\bibnamefont {Fredrick}}, \bibinfo {author}
  {\bibfnamefont {Q.}~\bibnamefont {Li}}, \bibinfo {author} {\bibfnamefont
  {D.}~\bibnamefont {Westly}}, \bibinfo {author} {\bibfnamefont {B.~R.}\
  \bibnamefont {Ilic}}, \bibinfo {author} {\bibfnamefont {B.}~\bibnamefont
  {Shen}}, \bibinfo {author} {\bibfnamefont {M.-G.}\ \bibnamefont {Suh}},
  \bibinfo {author} {\bibfnamefont {K.~Y.}\ \bibnamefont {Yang}}, \bibinfo
  {author} {\bibfnamefont {C.}~\bibnamefont {Johnson}}, \bibinfo {author}
  {\bibfnamefont {D.~M.~S.}\ \bibnamefont {Johnson}}, \bibinfo {author}
  {\bibfnamefont {L.}~\bibnamefont {Hollberg}}, \bibinfo {author}
  {\bibfnamefont {K.~J.}\ \bibnamefont {Vahala}}, \bibinfo {author}
  {\bibfnamefont {K.}~\bibnamefont {Srinivasan}}, \bibinfo {author}
  {\bibfnamefont {S.~A.}\ \bibnamefont {Diddams}}, \bibinfo {author}
  {\bibfnamefont {J.}~\bibnamefont {Kitching}}, \bibinfo {author}
  {\bibfnamefont {S.~B.}\ \bibnamefont {Papp}},\ and\ \bibinfo {author}
  {\bibfnamefont {M.~T.}\ \bibnamefont {Hummon}},\ }\bibfield  {title}
  {\bibinfo {title} {Architecture for the photonic integration of an optical
  atomic clock},\ }\href {https://doi.org/10.1364/OPTICA.6.000680} {\bibfield
  {journal} {\bibinfo  {journal} {Optica}\ }\textbf {\bibinfo {volume} {6}},\
  \bibinfo {pages} {680} (\bibinfo {year} {2019})}\BibitemShut {NoStop}%
\bibitem [{\citenamefont {Corcoran}\ \emph {et~al.}(2025)\citenamefont
  {Corcoran}, \citenamefont {Mitchell}, \citenamefont {Morandotti},
  \citenamefont {Oxenl{\o}we},\ and\ \citenamefont
  {Moss}}]{CorcoranNat.Photon.2025}%
  \BibitemOpen
  \bibfield  {author} {\bibinfo {author} {\bibfnamefont {B.}~\bibnamefont
  {Corcoran}}, \bibinfo {author} {\bibfnamefont {A.}~\bibnamefont {Mitchell}},
  \bibinfo {author} {\bibfnamefont {R.}~\bibnamefont {Morandotti}}, \bibinfo
  {author} {\bibfnamefont {L.~K.}\ \bibnamefont {Oxenl{\o}we}},\ and\ \bibinfo
  {author} {\bibfnamefont {D.~J.}\ \bibnamefont {Moss}},\ }\bibfield  {title}
  {\bibinfo {title} {Optical microcombs for ultrahigh-bandwidth
  communications},\ }\href {https://doi.org/10.1038/s41566-025-01662-9}
  {\bibfield  {journal} {\bibinfo  {journal} {Nature Photonics}\ }\textbf
  {\bibinfo {volume} {19}},\ \bibinfo {pages} {451} (\bibinfo {year}
  {2025})}\BibitemShut {NoStop}%
\bibitem [{\citenamefont {Dutt}\ \emph {et~al.}(2018)\citenamefont {Dutt},
  \citenamefont {Joshi}, \citenamefont {Ji}, \citenamefont {Cardenas},
  \citenamefont {Okawachi}, \citenamefont {Luke}, \citenamefont {Gaeta},\ and\
  \citenamefont {Lipson}}]{DuttSci.Adv.2018}%
  \BibitemOpen
  \bibfield  {author} {\bibinfo {author} {\bibfnamefont {A.}~\bibnamefont
  {Dutt}}, \bibinfo {author} {\bibfnamefont {C.}~\bibnamefont {Joshi}},
  \bibinfo {author} {\bibfnamefont {X.}~\bibnamefont {Ji}}, \bibinfo {author}
  {\bibfnamefont {J.}~\bibnamefont {Cardenas}}, \bibinfo {author}
  {\bibfnamefont {Y.}~\bibnamefont {Okawachi}}, \bibinfo {author}
  {\bibfnamefont {K.}~\bibnamefont {Luke}}, \bibinfo {author} {\bibfnamefont
  {A.~L.}\ \bibnamefont {Gaeta}},\ and\ \bibinfo {author} {\bibfnamefont
  {M.}~\bibnamefont {Lipson}},\ }\bibfield  {title} {\bibinfo {title} {On-chip
  dual-comb source for spectroscopy},\ }\href
  {https://doi.org/10.1126/sciadv.1701858} {\bibfield  {journal} {\bibinfo
  {journal} {Science Advances}\ }\textbf {\bibinfo {volume} {4}},\ \bibinfo
  {pages} {e1701858} (\bibinfo {year} {2018})}\BibitemShut {NoStop}%
\bibitem [{\citenamefont {Riemensberger}\ \emph {et~al.}(2020)\citenamefont
  {Riemensberger}, \citenamefont {Lukashchuk}, \citenamefont {Karpov},
  \citenamefont {Weng}, \citenamefont {Lucas}, \citenamefont {Liu},\ and\
  \citenamefont {Kippenberg}}]{RiemensbergerNature2020}%
  \BibitemOpen
  \bibfield  {author} {\bibinfo {author} {\bibfnamefont {J.}~\bibnamefont
  {Riemensberger}}, \bibinfo {author} {\bibfnamefont {A.}~\bibnamefont
  {Lukashchuk}}, \bibinfo {author} {\bibfnamefont {M.}~\bibnamefont {Karpov}},
  \bibinfo {author} {\bibfnamefont {W.}~\bibnamefont {Weng}}, \bibinfo {author}
  {\bibfnamefont {E.}~\bibnamefont {Lucas}}, \bibinfo {author} {\bibfnamefont
  {J.}~\bibnamefont {Liu}},\ and\ \bibinfo {author} {\bibfnamefont {T.~J.}\
  \bibnamefont {Kippenberg}},\ }\bibfield  {title} {\bibinfo {title} {Massively
  parallel coherent laser ranging using a soliton microcomb},\ }\href
  {https://doi.org/10.1038/s41586-020-2239-3} {\bibfield  {journal} {\bibinfo
  {journal} {Nature}\ }\textbf {\bibinfo {volume} {581}},\ \bibinfo {pages}
  {164} (\bibinfo {year} {2020})}\BibitemShut {NoStop}%
\bibitem [{\citenamefont {Huang}\ \emph {et~al.}(2019)\citenamefont {Huang},
  \citenamefont {Lucas}, \citenamefont {Liu}, \citenamefont {Raja},
  \citenamefont {Lihachev}, \citenamefont {Gorodetsky}, \citenamefont
  {Engelsen},\ and\ \citenamefont {Kippenberg}}]{HuangPhys.Rev.A2019}%
  \BibitemOpen
  \bibfield  {author} {\bibinfo {author} {\bibfnamefont {G.}~\bibnamefont
  {Huang}}, \bibinfo {author} {\bibfnamefont {E.}~\bibnamefont {Lucas}},
  \bibinfo {author} {\bibfnamefont {J.}~\bibnamefont {Liu}}, \bibinfo {author}
  {\bibfnamefont {A.~S.}\ \bibnamefont {Raja}}, \bibinfo {author}
  {\bibfnamefont {G.}~\bibnamefont {Lihachev}}, \bibinfo {author}
  {\bibfnamefont {M.~L.}\ \bibnamefont {Gorodetsky}}, \bibinfo {author}
  {\bibfnamefont {N.~J.}\ \bibnamefont {Engelsen}},\ and\ \bibinfo {author}
  {\bibfnamefont {T.~J.}\ \bibnamefont {Kippenberg}},\ }\bibfield  {title}
  {\bibinfo {title} {Thermorefractive noise in silicon-nitride
  microresonators},\ }\href {https://doi.org/10.1103/PhysRevA.99.061801}
  {\bibfield  {journal} {\bibinfo  {journal} {Physical Review A}\ }\textbf
  {\bibinfo {volume} {99}},\ \bibinfo {pages} {061801} (\bibinfo {year}
  {2019})}\BibitemShut {NoStop}%
\bibitem [{\citenamefont {Drake}\ \emph {et~al.}(2020)\citenamefont {Drake},
  \citenamefont {Stone}, \citenamefont {Briles},\ and\ \citenamefont
  {Papp}}]{DrakeNat.Photonics2020}%
  \BibitemOpen
  \bibfield  {author} {\bibinfo {author} {\bibfnamefont {T.~E.}\ \bibnamefont
  {Drake}}, \bibinfo {author} {\bibfnamefont {J.~R.}\ \bibnamefont {Stone}},
  \bibinfo {author} {\bibfnamefont {T.~C.}\ \bibnamefont {Briles}},\ and\
  \bibinfo {author} {\bibfnamefont {S.~B.}\ \bibnamefont {Papp}},\ }\bibfield
  {title} {\bibinfo {title} {Thermal decoherence and laser cooling of {{Kerr}}
  microresonator solitons},\ }\href {https://doi.org/10.1038/s41566-020-0651-8}
  {\bibfield  {journal} {\bibinfo  {journal} {Nature Photonics}\ }\textbf
  {\bibinfo {volume} {14}},\ \bibinfo {pages} {480} (\bibinfo {year}
  {2020})}\BibitemShut {NoStop}%
\bibitem [{\citenamefont {Telle}\ \emph {et~al.}(1999)\citenamefont {Telle},
  \citenamefont {Steinmeyer}, \citenamefont {Dunlop}, \citenamefont {Stenger},
  \citenamefont {Sutter},\ and\ \citenamefont {Keller}}]{TelleApplPhysB1999a}%
  \BibitemOpen
  \bibfield  {author} {\bibinfo {author} {\bibfnamefont {H.}~\bibnamefont
  {Telle}}, \bibinfo {author} {\bibfnamefont {G.}~\bibnamefont {Steinmeyer}},
  \bibinfo {author} {\bibfnamefont {A.}~\bibnamefont {Dunlop}}, \bibinfo
  {author} {\bibfnamefont {J.}~\bibnamefont {Stenger}}, \bibinfo {author}
  {\bibfnamefont {D.}~\bibnamefont {Sutter}},\ and\ \bibinfo {author}
  {\bibfnamefont {U.}~\bibnamefont {Keller}},\ }\bibfield  {title} {\bibinfo
  {title} {Carrier-envelope offset phase control: {{A}} novel concept for
  absolute optical frequency measurement and ultrashort pulse generation},\
  }\href {https://doi.org/10.1007/s003400050813} {\bibfield  {journal}
  {\bibinfo  {journal} {Applied Physics B}\ }\textbf {\bibinfo {volume} {69}},\
  \bibinfo {pages} {327} (\bibinfo {year} {1999})}\BibitemShut {NoStop}%
\bibitem [{\citenamefont {Moille}\ \emph {et~al.}(2023)\citenamefont {Moille},
  \citenamefont {Stone}, \citenamefont {Chojnacky}, \citenamefont {Shrestha},
  \citenamefont {Javid}, \citenamefont {Menyuk},\ and\ \citenamefont
  {Srinivasan}}]{MoilleNature2023}%
  \BibitemOpen
  \bibfield  {author} {\bibinfo {author} {\bibfnamefont {G.}~\bibnamefont
  {Moille}}, \bibinfo {author} {\bibfnamefont {J.}~\bibnamefont {Stone}},
  \bibinfo {author} {\bibfnamefont {M.}~\bibnamefont {Chojnacky}}, \bibinfo
  {author} {\bibfnamefont {R.}~\bibnamefont {Shrestha}}, \bibinfo {author}
  {\bibfnamefont {U.~A.}\ \bibnamefont {Javid}}, \bibinfo {author}
  {\bibfnamefont {C.}~\bibnamefont {Menyuk}},\ and\ \bibinfo {author}
  {\bibfnamefont {K.}~\bibnamefont {Srinivasan}},\ }\bibfield  {title}
  {\bibinfo {title} {Kerr-induced synchronization of a cavity soliton to an
  optical reference},\ }\href {https://doi.org/10.1038/s41586-023-06730-0}
  {\bibfield  {journal} {\bibinfo  {journal} {Nature}\ }\textbf {\bibinfo
  {volume} {624}},\ \bibinfo {pages} {267} (\bibinfo {year}
  {2023})}\BibitemShut {NoStop}%
\bibitem [{\citenamefont {Moille}\ \emph
  {et~al.}(2025{\natexlab{a}})\citenamefont {Moille}, \citenamefont {Sridhar},
  \citenamefont {Shandilya}, \citenamefont {Dutt}, \citenamefont {Menyuk},\
  and\ \citenamefont {Srinivasan}}]{MoillePhys.Rev.Lett.2025b}%
  \BibitemOpen
  \bibfield  {author} {\bibinfo {author} {\bibfnamefont {G.}~\bibnamefont
  {Moille}}, \bibinfo {author} {\bibfnamefont {S.~K.}\ \bibnamefont {Sridhar}},
  \bibinfo {author} {\bibfnamefont {P.}~\bibnamefont {Shandilya}}, \bibinfo
  {author} {\bibfnamefont {A.}~\bibnamefont {Dutt}}, \bibinfo {author}
  {\bibfnamefont {C.}~\bibnamefont {Menyuk}},\ and\ \bibinfo {author}
  {\bibfnamefont {K.}~\bibnamefont {Srinivasan}},\ }\bibfield  {title}
  {\bibinfo {title} {Toward {{Chaotic Group Velocity Hopping}} of an {{On-Chip
  Dissipative Kerr Soliton}}},\ }\href {https://doi.org/10.1103/2k7d-p7rm}
  {\bibfield  {journal} {\bibinfo  {journal} {Physical Review Letters}\
  }\textbf {\bibinfo {volume} {135}},\ \bibinfo {pages} {133802} (\bibinfo
  {year} {2025}{\natexlab{a}})}\BibitemShut {NoStop}%
\bibitem [{\citenamefont {Sun}\ \emph {et~al.}(2025)\citenamefont {Sun},
  \citenamefont {Harrington}, \citenamefont {Tabatabaei}, \citenamefont
  {Hanifi}, \citenamefont {Liu}, \citenamefont {Wang}, \citenamefont {Wang},
  \citenamefont {Yang}, \citenamefont {Liu}, \citenamefont {Morgan},
  \citenamefont {Bowers}, \citenamefont {Morton}, \citenamefont {Nelson},
  \citenamefont {Beling}, \citenamefont {Blumenthal},\ and\ \citenamefont
  {Yi}}]{SunNat.Photon.2025}%
  \BibitemOpen
  \bibfield  {author} {\bibinfo {author} {\bibfnamefont {S.}~\bibnamefont
  {Sun}}, \bibinfo {author} {\bibfnamefont {M.~W.}\ \bibnamefont {Harrington}},
  \bibinfo {author} {\bibfnamefont {F.}~\bibnamefont {Tabatabaei}}, \bibinfo
  {author} {\bibfnamefont {S.}~\bibnamefont {Hanifi}}, \bibinfo {author}
  {\bibfnamefont {K.}~\bibnamefont {Liu}}, \bibinfo {author} {\bibfnamefont
  {J.}~\bibnamefont {Wang}}, \bibinfo {author} {\bibfnamefont {B.}~\bibnamefont
  {Wang}}, \bibinfo {author} {\bibfnamefont {Z.}~\bibnamefont {Yang}}, \bibinfo
  {author} {\bibfnamefont {R.}~\bibnamefont {Liu}}, \bibinfo {author}
  {\bibfnamefont {J.~S.}\ \bibnamefont {Morgan}}, \bibinfo {author}
  {\bibfnamefont {S.~M.}\ \bibnamefont {Bowers}}, \bibinfo {author}
  {\bibfnamefont {P.~A.}\ \bibnamefont {Morton}}, \bibinfo {author}
  {\bibfnamefont {K.~D.}\ \bibnamefont {Nelson}}, \bibinfo {author}
  {\bibfnamefont {A.}~\bibnamefont {Beling}}, \bibinfo {author} {\bibfnamefont
  {D.~J.}\ \bibnamefont {Blumenthal}},\ and\ \bibinfo {author} {\bibfnamefont
  {X.}~\bibnamefont {Yi}},\ }\bibfield  {title} {\bibinfo {title} {Microcavity
  {{Kerr}} optical frequency division with integrated {{SiN}} photonics},\
  }\href {https://doi.org/10.1038/s41566-025-01668-3} {\bibfield  {journal}
  {\bibinfo  {journal} {Nature Photonics}\ ,\ \bibinfo {pages} {1}} (\bibinfo
  {year} {2025})}\BibitemShut {NoStop}%
\bibitem [{\citenamefont {Moille}\ \emph
  {et~al.}(2025{\natexlab{b}})\citenamefont {Moille}, \citenamefont
  {Shandilya}, \citenamefont {Stone}, \citenamefont {Menyuk},\ and\
  \citenamefont {Srinivasan}}]{MoilleOptica2025}%
  \BibitemOpen
  \bibfield  {author} {\bibinfo {author} {\bibfnamefont {G.}~\bibnamefont
  {Moille}}, \bibinfo {author} {\bibfnamefont {P.}~\bibnamefont {Shandilya}},
  \bibinfo {author} {\bibfnamefont {J.}~\bibnamefont {Stone}}, \bibinfo
  {author} {\bibfnamefont {C.}~\bibnamefont {Menyuk}},\ and\ \bibinfo {author}
  {\bibfnamefont {K.}~\bibnamefont {Srinivasan}},\ }\bibfield  {title}
  {\bibinfo {title} {All-optical noise quenching of an integrated frequency
  comb},\ }\href {https://doi.org/10.1364/OPTICA.561954} {\bibfield  {journal}
  {\bibinfo  {journal} {Optica}\ }\textbf {\bibinfo {volume} {12}},\ \bibinfo
  {pages} {1020} (\bibinfo {year} {2025}{\natexlab{b}})}\BibitemShut {NoStop}%
\bibitem [{\citenamefont {Taheri}\ \emph {et~al.}(2017)\citenamefont {Taheri},
  \citenamefont {Matsko},\ and\ \citenamefont
  {Maleki}}]{TaheriEur.Phys.J.D2017}%
  \BibitemOpen
  \bibfield  {author} {\bibinfo {author} {\bibfnamefont {H.}~\bibnamefont
  {Taheri}}, \bibinfo {author} {\bibfnamefont {A.~B.}\ \bibnamefont {Matsko}},\
  and\ \bibinfo {author} {\bibfnamefont {L.}~\bibnamefont {Maleki}},\
  }\bibfield  {title} {\bibinfo {title} {Optical lattice trap for {{Kerr}}
  solitons},\ }\href {https://doi.org/10.1140/epjd/e2017-80150-6} {\bibfield
  {journal} {\bibinfo  {journal} {The European Physical Journal D}\ }\textbf
  {\bibinfo {volume} {71}},\ \bibinfo {pages} {153} (\bibinfo {year}
  {2017})}\BibitemShut {NoStop}%
\bibitem [{\citenamefont {Kaup}()}]{KaupPhysRevA1990}%
  \BibitemOpen
  \bibfield  {author} {\bibinfo {author} {\bibfnamefont {D.~J.}\ \bibnamefont
  {Kaup}},\ }\bibfield  {title} {\bibinfo {title} {Perturbation theory for
  solitons in optical fibers},\ }\href
  {https://doi.org/10.1103/PhysRevA.42.5689} {\ \textbf {\bibinfo {volume}
  {42}},\ \bibinfo {pages} {5689}}\BibitemShut {NoStop}%
\bibitem [{\citenamefont {Haus}\ and\ \citenamefont {Lai}(1990)}]{Haus1990}%
  \BibitemOpen
  \bibfield  {author} {\bibinfo {author} {\bibfnamefont {H.~A.}\ \bibnamefont
  {Haus}}\ and\ \bibinfo {author} {\bibfnamefont {Y.}~\bibnamefont {Lai}},\
  }\bibfield  {title} {\bibinfo {title} {Quantum theory of soliton squeezing: a
  linearized approach},\ }\href {https://doi.org/10.1364/JOSAB.7.000386}
  {\bibfield  {journal} {\bibinfo  {journal} {J. Opt. Soc. Am. B}\ }\textbf
  {\bibinfo {volume} {7}},\ \bibinfo {pages} {386} (\bibinfo {year}
  {1990})}\BibitemShut {NoStop}%
\bibitem [{\citenamefont {Georges}()}]{GeorgesOptFibTech1995}%
  \BibitemOpen
  \bibfield  {author} {\bibinfo {author} {\bibfnamefont {T.}~\bibnamefont
  {Georges}},\ }\bibfield  {title} {\bibinfo {title} {Perturbation {{Theory}}
  for the {{Assessment}} of {{Soliton Transmission Control}}},\ }\href
  {https://doi.org/10.1006/ofte.1995.1001} {\ \textbf {\bibinfo {volume} {1}},\
  \bibinfo {pages} {97}}\BibitemShut {NoStop}%
\bibitem [{\citenamefont {Mizrahi}\ \emph {et~al.}(2024)\citenamefont
  {Mizrahi}, \citenamefont {Courtright}, \citenamefont {Shandilya},
  \citenamefont {Menyuk},\ and\ \citenamefont {Gat}}]{Mizrahi2024}%
  \BibitemOpen
  \bibfield  {author} {\bibinfo {author} {\bibfnamefont {J.~P.}\ \bibnamefont
  {Mizrahi}}, \bibinfo {author} {\bibfnamefont {L.}~\bibnamefont {Courtright}},
  \bibinfo {author} {\bibfnamefont {P.}~\bibnamefont {Shandilya}}, \bibinfo
  {author} {\bibfnamefont {C.~R.}\ \bibnamefont {Menyuk}},\ and\ \bibinfo
  {author} {\bibfnamefont {O.}~\bibnamefont {Gat}},\ }\bibfield  {title}
  {\bibinfo {title} {Soliton synchronization in microresonators with a
  modulated pump},\ }\href {https://doi.org/10.1103/PhysRevE.109.064204}
  {\bibfield  {journal} {\bibinfo  {journal} {Phys. Rev. E}\ }\textbf {\bibinfo
  {volume} {109}},\ \bibinfo {pages} {064204} (\bibinfo {year}
  {2024})}\BibitemShut {NoStop}%
\bibitem [{\citenamefont {Leshem}\ \emph {et~al.}(2025)\citenamefont {Leshem},
  \citenamefont {Akter}, \citenamefont {Courtright}, \citenamefont {Shandilya},
  \citenamefont {Menyuk},\ and\ \citenamefont {Gat}}]{Leshem2025}%
  \BibitemOpen
  \bibfield  {author} {\bibinfo {author} {\bibfnamefont {A.}~\bibnamefont
  {Leshem}}, \bibinfo {author} {\bibfnamefont {S.}~\bibnamefont {Akter}},
  \bibinfo {author} {\bibfnamefont {L.}~\bibnamefont {Courtright}}, \bibinfo
  {author} {\bibfnamefont {P.}~\bibnamefont {Shandilya}}, \bibinfo {author}
  {\bibfnamefont {C.~R.}\ \bibnamefont {Menyuk}},\ and\ \bibinfo {author}
  {\bibfnamefont {O.}~\bibnamefont {Gat}},\ }\href
  {https://arxiv.org/abs/2507.07851} {\bibinfo {title} {Dynamics of interacting
  cavity solitons}} (\bibinfo {year} {2025}),\ \Eprint
  {https://arxiv.org/abs/2507.07851} {arXiv:2507.07851 [nlin.PS]} \BibitemShut
  {NoStop}%
\bibitem [{\citenamefont {Wildi}\ \emph {et~al.}(2023)\citenamefont {Wildi},
  \citenamefont {Ulanov}, \citenamefont {Englebert}, \citenamefont {Voumard},\
  and\ \citenamefont {Herr}}]{WildiAPLPhotonics2023}%
  \BibitemOpen
  \bibfield  {author} {\bibinfo {author} {\bibfnamefont {T.}~\bibnamefont
  {Wildi}}, \bibinfo {author} {\bibfnamefont {A.}~\bibnamefont {Ulanov}},
  \bibinfo {author} {\bibfnamefont {N.}~\bibnamefont {Englebert}}, \bibinfo
  {author} {\bibfnamefont {T.}~\bibnamefont {Voumard}},\ and\ \bibinfo {author}
  {\bibfnamefont {T.}~\bibnamefont {Herr}},\ }\bibfield  {title} {\bibinfo
  {title} {Sideband injection locking in microresonator frequency combs},\
  }\href {https://doi.org/10.1063/5.0170224} {\bibfield  {journal} {\bibinfo
  {journal} {APL Photonics}\ }\textbf {\bibinfo {volume} {8}},\ \bibinfo
  {pages} {120801} (\bibinfo {year} {2023})}\BibitemShut {NoStop}%
\bibitem [{\citenamefont {Englebert}\ \emph {et~al.}(2024)\citenamefont
  {Englebert}, \citenamefont {Simon}, \citenamefont {Arab{\'i}}, \citenamefont
  {Leo},\ and\ \citenamefont {Gorza}}]{Englebert2024}%
  \BibitemOpen
  \bibfield  {author} {\bibinfo {author} {\bibfnamefont {N.}~\bibnamefont
  {Englebert}}, \bibinfo {author} {\bibfnamefont {C.}~\bibnamefont {Simon}},
  \bibinfo {author} {\bibfnamefont {C.~M.}\ \bibnamefont {Arab{\'i}}}, \bibinfo
  {author} {\bibfnamefont {F.}~\bibnamefont {Leo}},\ and\ \bibinfo {author}
  {\bibfnamefont {S.-P.}\ \bibnamefont {Gorza}},\ }\href
  {https://doi.org/10.48550/arXiv.2406.12848} {\bibinfo {title} {Manipulation
  and control of temporal cavity solitons with trapping potential}} (\bibinfo
  {year} {2024}),\ \Eprint {https://arxiv.org/abs/2406.12848} {arXiv:2406.12848
  [physics]} \BibitemShut {NoStop}%
\bibitem [{\citenamefont {Mili\'an}\ \emph {et~al.}(2015)\citenamefont
  {Mili\'an}, \citenamefont {Gorbach}, \citenamefont {Taki}, \citenamefont
  {Yulin},\ and\ \citenamefont {Skryabin}}]{MilianPhysRevA2015}%
  \BibitemOpen
  \bibfield  {author} {\bibinfo {author} {\bibfnamefont {C.}~\bibnamefont
  {Mili\'an}}, \bibinfo {author} {\bibfnamefont {A.~V.}\ \bibnamefont
  {Gorbach}}, \bibinfo {author} {\bibfnamefont {M.}~\bibnamefont {Taki}},
  \bibinfo {author} {\bibfnamefont {A.~V.}\ \bibnamefont {Yulin}},\ and\
  \bibinfo {author} {\bibfnamefont {D.~V.}\ \bibnamefont {Skryabin}},\
  }\bibfield  {title} {\bibinfo {title} {Solitons and frequency combs in silica
  microring resonators: Interplay of the raman and higher-order dispersion
  effects},\ }\href {https://doi.org/10.1103/PhysRevA.92.033851} {\bibfield
  {journal} {\bibinfo  {journal} {Phys. Rev. A}\ }\textbf {\bibinfo {volume}
  {92}},\ \bibinfo {pages} {033851} (\bibinfo {year} {2015})}\BibitemShut
  {NoStop}%
\bibitem [{\citenamefont {Yi}\ \emph {et~al.}()\citenamefont {Yi},
  \citenamefont {Yang}, \citenamefont {Yang},\ and\ \citenamefont
  {Vahala}}]{YiOptLett2016}%
  \BibitemOpen
  \bibfield  {author} {\bibinfo {author} {\bibfnamefont {X.}~\bibnamefont
  {Yi}}, \bibinfo {author} {\bibfnamefont {Q.-F.}\ \bibnamefont {Yang}},
  \bibinfo {author} {\bibfnamefont {K.~Y.}\ \bibnamefont {Yang}},\ and\
  \bibinfo {author} {\bibfnamefont {K.}~\bibnamefont {Vahala}},\ }\bibfield
  {title} {\bibinfo {title} {Theory and measurement of the soliton
  self-frequency shift and efficiency in optical microcavities},\ }\href
  {https://doi.org/10.1364/OL.41.003419} {\ \textbf {\bibinfo {volume} {41}},\
  \bibinfo {pages} {3419}}\BibitemShut {NoStop}%
\bibitem [{\citenamefont {Cherenkov}\ \emph {et~al.}(2017)\citenamefont
  {Cherenkov}, \citenamefont {Lobanov},\ and\ \citenamefont
  {Gorodetsky}}]{CherenkovPhysRevA2017}%
  \BibitemOpen
  \bibfield  {author} {\bibinfo {author} {\bibfnamefont {A.~V.}\ \bibnamefont
  {Cherenkov}}, \bibinfo {author} {\bibfnamefont {V.~E.}\ \bibnamefont
  {Lobanov}},\ and\ \bibinfo {author} {\bibfnamefont {M.~L.}\ \bibnamefont
  {Gorodetsky}},\ }\bibfield  {title} {\bibinfo {title} {Dissipative kerr
  solitons and cherenkov radiation in optical microresonators with third-order
  dispersion},\ }\href {https://doi.org/10.1103/PhysRevA.95.033810} {\bibfield
  {journal} {\bibinfo  {journal} {Phys. Rev. A}\ }\textbf {\bibinfo {volume}
  {95}},\ \bibinfo {pages} {033810} (\bibinfo {year} {2017})}\BibitemShut
  {NoStop}%
\bibitem [{\citenamefont {Akhmediev}\ and\ \citenamefont
  {Karlsson}(1995)}]{akhmediev_cherenkov_1995}%
  \BibitemOpen
  \bibfield  {author} {\bibinfo {author} {\bibfnamefont {N.}~\bibnamefont
  {Akhmediev}}\ and\ \bibinfo {author} {\bibfnamefont {M.}~\bibnamefont
  {Karlsson}},\ }\bibfield  {title} {\bibinfo {title} {Cherenkov radiation
  emitted by solitons in optical fibers},\ }\href
  {https://doi.org/10.1103/PhysRevA.51.2602} {\bibfield  {journal} {\bibinfo
  {journal} {Physical Review A}\ }\textbf {\bibinfo {volume} {51}},\ \bibinfo
  {pages} {2602} (\bibinfo {year} {1995})}\BibitemShut {NoStop}%
\bibitem [{\citenamefont {Kodama}\ and\ \citenamefont
  {Hasegawa}(1987)}]{kodama_nonlinear_1987}%
  \BibitemOpen
  \bibfield  {author} {\bibinfo {author} {\bibfnamefont {Y.}~\bibnamefont
  {Kodama}}\ and\ \bibinfo {author} {\bibfnamefont {A.}~\bibnamefont
  {Hasegawa}},\ }\bibfield  {title} {\bibinfo {title} {Nonlinear pulse
  propagation in a monomode dielectric guide},\ }\href
  {https://doi.org/10.1109/JQE.1987.1073392} {\bibfield  {journal} {\bibinfo
  {journal} {IEEE Journal of Quantum Electronics}\ }\textbf {\bibinfo {volume}
  {23}},\ \bibinfo {pages} {510} (\bibinfo {year} {1987})}\BibitemShut
  {NoStop}%
\bibitem [{\citenamefont {Zhou}\ \emph {et~al.}(2019)\citenamefont {Zhou},
  \citenamefont {Geng}, \citenamefont {Cui}, \citenamefont {Huang},
  \citenamefont {Zhou}, \citenamefont {Qiu},\ and\ \citenamefont
  {Wei~Wong}}]{ZhouLightSciAppl2019}%
  \BibitemOpen
  \bibfield  {author} {\bibinfo {author} {\bibfnamefont {H.}~\bibnamefont
  {Zhou}}, \bibinfo {author} {\bibfnamefont {Y.}~\bibnamefont {Geng}}, \bibinfo
  {author} {\bibfnamefont {W.}~\bibnamefont {Cui}}, \bibinfo {author}
  {\bibfnamefont {S.-W.}\ \bibnamefont {Huang}}, \bibinfo {author}
  {\bibfnamefont {Q.}~\bibnamefont {Zhou}}, \bibinfo {author} {\bibfnamefont
  {K.}~\bibnamefont {Qiu}},\ and\ \bibinfo {author} {\bibfnamefont
  {C.}~\bibnamefont {Wei~Wong}},\ }\bibfield  {title} {\bibinfo {title}
  {Soliton bursts and deterministic dissipative {{Kerr}} soliton generation in
  auxiliary-assisted microcavities},\ }\href
  {https://doi.org/10.1038/s41377-019-0161-y} {\bibfield  {journal} {\bibinfo
  {journal} {Light: Science \& Applications}\ }\textbf {\bibinfo {volume}
  {8}},\ \bibinfo {pages} {50} (\bibinfo {year} {2019})}\BibitemShut {NoStop}%
\bibitem [{\citenamefont {Zhang}\ \emph {et~al.}(2019)\citenamefont {Zhang},
  \citenamefont {Silver}, \citenamefont {Del~Bino}, \citenamefont {Copie},
  \citenamefont {Woodley}, \citenamefont {Ghalanos}, \citenamefont {Svela},
  \citenamefont {Moroney},\ and\ \citenamefont {Del'Haye}}]{ZhangOptica2019}%
  \BibitemOpen
  \bibfield  {author} {\bibinfo {author} {\bibfnamefont {S.}~\bibnamefont
  {Zhang}}, \bibinfo {author} {\bibfnamefont {J.~M.}\ \bibnamefont {Silver}},
  \bibinfo {author} {\bibfnamefont {L.}~\bibnamefont {Del~Bino}}, \bibinfo
  {author} {\bibfnamefont {F.}~\bibnamefont {Copie}}, \bibinfo {author}
  {\bibfnamefont {M.~T.~M.}\ \bibnamefont {Woodley}}, \bibinfo {author}
  {\bibfnamefont {G.~N.}\ \bibnamefont {Ghalanos}}, \bibinfo {author}
  {\bibfnamefont {A.~{\O}.}\ \bibnamefont {Svela}}, \bibinfo {author}
  {\bibfnamefont {N.}~\bibnamefont {Moroney}},\ and\ \bibinfo {author}
  {\bibfnamefont {P.}~\bibnamefont {Del'Haye}},\ }\bibfield  {title} {\bibinfo
  {title} {Sub-milliwatt-level microresonator solitons with extended access
  range using an auxiliary laser},\ }\href
  {https://doi.org/10.1364/OPTICA.6.000206} {\bibfield  {journal} {\bibinfo
  {journal} {Optica}\ }\textbf {\bibinfo {volume} {6}},\ \bibinfo {pages} {206}
  (\bibinfo {year} {2019})}\BibitemShut {NoStop}%
\end{thebibliography}
